\documentclass[fleqn,usenatbib]{mnras}

\usepackage{newtxtext,newtxmath}
\usepackage{xcolor}

\usepackage[T1]{fontenc}

\DeclareRobustCommand{\VAN}[3]{#2}
\let\VANthebibliography\thebibliography
\def\thebibliography{\DeclareRobustCommand{\VAN}[3]{##3}\VANthebibliography}

\usepackage{graphicx}	
\usepackage{amsmath}	

\title[Optical Emission from BBH remnants in AGNs ]{
Optical Emission Model for Binary Black Hole
Merger Remnants Travelling through Discs of Active Galactic Nuclei
}

\author[J. C. Rodr\'iguez-Ram\'irez]{
J. C. Rodr\'iguez-Ram\'irez,$^{1}$\thanks{E-mail: juancr@cbpf.br}
C. R. Bom$^{1,2}$, B. Fraga$^{1}$, R. Nemmen$^{3,4}$
\\
$^{1}$ Centro Brasileiro de Pesquisas Fisicas (CBPF), 
Rua Dr Xavier Sigaud 150, CEP 22290-180 Rio de Janeiro RJ, Brazil\\
$^{2}$Centro Federal de Educa\c{c}\~{a}o Tecnol\'{o}gica Celso Suckow da Fonseca, Rodovia M\'{a}rcio Covas, lote J2, quadra J - Itagua\'{i} (Brazil) \\
$^{3}$Instituto de Astronomia, Geof\'{\i}sica e Ci\^encias Atmosf\'ericas, Universidade de S\~ao Paulo, S\~ao Paulo, SP, 05508-090, Brazil\\
$^4$Kavli Institute for Particle Astrophysics and Cosmology, Stanford University, Stanford, CA 94305, USA 
}

\pubyear{2023}

\begin{document}
\label{firstpage}
\pagerange{\pageref{firstpage}--\pageref{lastpage}}
\maketitle

\begin{abstract}
Active galactic nuclei (AGNs) have been proposed as plausible sites for hosting a sizable 
fraction of the binary black hole (BBH) mergers measured through gravitational waves 
(GWs) by the LIGO-Virgo-Kagra (LVK) experiment.     
These GWs could be accompanied by radiation feedback due to the interaction of the BBH merger
remnant with the AGN disc.
We present a new predicted radiation signature
driven by the passage
of a kicked BBH remnant throughout a thin AGN disc.
We analyse the situation of a merger occurring outside the thin disc, 
where the merger is of  second or higher generation in a merging hierarchical sequence. 
The coalescence produces a kicked BH remnant that eventually plunges into the disc, accretes material, and inflates jet cocoons.
We consider the case of a jet cocoon propagating quasi-parallel to the disc plane
and study the outflow that results when the cocoon emerges from the disc.
We calculate the transient emission of the emerging cocoon
using a photon diffusion model typically employed to describe the light curves of supernovae.
Depending on the parameter configuration, the flare produced by the
emerging cocoon
could be comparable to or exceed the AGN background emission at optical, and extreme ultraviolet wavelengths.
For instance, in AGNs with 
central engines of
$\sim 5\times10^{6}$ M$_\odot$, flares driven by BH remnants
with masses of $\sim$ 100 M$_\odot$
can appear in about $\sim$[10-100] days after the GW, lasting for few days.
\end{abstract}




\section{Introduction}

A natural prediction of stellar dynamics is the clustering of 
thousands of stellar mass black holes (sBHs) orbiting a few parsecs
around the central supermassive  black holes (SMBHs) of galaxies
\citep{1977ApJ...216..883B, 1993ApJ...408..496M, 2000ApJ...545..847M, 2018Natur.556...70H}.
In active galactic nuclei (AGNs), these sBHs can receive torque forces bringing their orbits aligned to
the plane of the AGN disc
\citep{2012MNRAS.425..460M,2018ApJ...866...66M},
and migrate into certain radial regions,
denoted as migration traps 
\citep{2016ApJ...819L..17B, 2019ApJ...878...85S}.
Under this scenario, such migrations induce close 
encounters among compact objects, thus making  AGNs discs
plausible candidates for hosting the
binary black holes (BBH)
coalescences detected
by the LIGO-Virgo-Kagra (LVK) experiment
\citep{2021arXiv211103606T, 2017NatCo...8..831B}.

Remnants from non-symmetrical binary  mergers
can be born with recoil or kick velocities of
$\sim$[100 - 1000] 
Km s$^{-1}$
(\citealt{2007PhRvL..98w1102C, 2007ApJ...659L...5C, 2015PhRvD..92b4022Z,2022PhRvL.128s1102V}).
In typical AGNs, these kicked remnants can be retained by 
the AGN potential well,  having then the chance to undergo further encounters with compact objects.
Thus,
AGNs could also host BH hierarchical growth,
i.e., BHs growing through binary mergers where one or both
components are the remnants of a previous coalescence.
Hierarchical merging in AGNs is a viable channel to explain 
the observed anti-correlation among the mass ratio and effective spin of BH mergers \citep{PhysRevD.108.083033} as well as 
mergers with components of masses $\gtrsim$ 50 M$_\odot$ 
\citep{2019PhRvD.100j4015C, 2020RNAAS...4....2K,2020PhRvL.125j1102A},
whose existence challenges standard stellar evolution models \citep{2017ApJ...836..244W,2020ApJ...900L..13A}.
Hierarchical growth is also a possible channel for the formation of
intermediate-mass black holes (IMBHs)
\citep{2012MNRAS.425..460M,2014MNRAS.441..900M}, 
a class of BHs with mass in the range [10$^2$ - 10$^5$] M$_\odot$,
thought to be the seed of SMBHs, and showing currently
poor observational evidence compared to their stellar mass and super-massive counterparts
\citep{2020ARA&A..58..257G}.

An appealing property of the AGN merger channel, 
not present in other BBH merger scenarios,
is the ability to generate multimessenger emissions.
Contrary to gravitational waves from compact binaries 
involving neutron stars, BBH mergers are not expected to produce electromagnetic (EM) counterparts by themselves.
However, BBHs coalescing nearby high-dense media, like the thin discs of AGNs, may produce
multimessenger emission (e.g., GWs and radiation) due to the interaction of the binary and/or the remnant with the  gas of 
the disc
\citep{2019ApJ...884L..50M,2020PhRvL.124y1102G,2021ApJ...916..111K,2021ApJ...916L..17W, 2023ApJ...950...13T,2023arXiv231018392T}.

Among the GWs measured by the LIGO-Virgo experiment, 
the event GW190521 has been of particular interest 
in the context of mergers assisted by AGNs.
This GW event produced a BH remnant of $\sim$ 142 M$_\odot$, 
the heaviest BH measured by an LVK observation so far,
from a merger with inferred components of 85 and 55 M$_\odot$
\citep{2020PhRvL.125j1102A}.
Such massive binary components are unlikely of stellar collapse origin 
\citep{2017ApJ...836..244W}
and more
naturally explained by the hierarchical merger channel.
Moreover, an optical flare in the AGN J1249 + 3449 at redshift $z=0.438$ was
claimed  EM counterpart to GW190521 by \cite{2020PhRvL.124y1102G},
who interpreted the EM flare as originated in a hyper-accretion episode onto 
the merger remnant.
This GW event has one of the largest localisation volumes on the
sky among the events
measured by LVK so far \citep{2021ApJ...914L..34P}, which makes the multi-messenger
association a subject of ongoing debate
\citep{2021CQGra..38w5004A,
2021ApJ...914L..34P,
2022MNRAS.513.2152C,2020MNRAS.499L..87D}.
If established, the aforementioned
and future  
EM counterparts to GWs can serve as direct probes of 
AGN discs, 
could
be employed to derive cosmological constraints (e.g., \citealt{2020RNAAS...4..209H} ),
as well as an independent method for measuring the Hubble constant \citep{2021ApJ...909..218A,2020arXiv200914247G,2023arXiv230701330B}.
Further theoretical and observational analyses are then
timely to solidly localise EM signals associated with GW events.


This paper presents a new predicted EM signature triggered by 
BBH mergers in AGNs.
We consider a BBH merger occurring outside and nearby the plane of an AGN thin disc,
whose kicked remnant eventually plunges and traverses the disc.
We study one of the jet cocoons
\citep{2011ApJ...740..100B}
driven by the remnant within the disc,
focusing on the case when the cocoon propagates quasi-parallel to the disc.
The jet cocoon eventually emerges
outside the disc in the form of a non-relativistic outflow.
We then calculate the emission produced after the
emerging cocoon expands and let the thermal photons escape.
The proposed emission scenario is motivated by previous works that consider the extraction of
disc material as a viable mechanism to explain thermal and non-thermal flares from AGNs 
\citep{1998ApJ...507..131I,2016MNRAS.457.1145P,2019ApJ...882...88V,2020MNRAS.498.5424R}.

The present EM counterpart scenario is also motivated by second or 
higher-generation mergers in a hierarchical sequence which could explain 
mergers detected by LVK with components more massive than predicted by 
stellar evolution models (e.g., GW190521
\citealt{2020ApJ...900L..13A} 
and GW170729
\citealt{PhysRevX.9.031040}). 
The components of such binary systems have then been born in previous 
coalescences whose recoils perturb the alignment of the remnants with
the disc plane.
Thus, mergers of second or higher generations could occur within or
outside the disc and here we consider the latter case (different to the 
previous works of
\citealt{2019ApJ...884L..50M,2020PhRvL.124y1102G,2021ApJ...916..111K,2021ApJ...916L..17W,
2023ApJ...942...99G,
2023ApJ...950...13T},
that consider mergers 
within the disc).

The paper is organised as follows.
Section~\ref{sec:model} describes the proposed emission scenario and derives an analytic model for the emitted spectrum.  
In Section~\ref{sec:profiles}, we discuss the parametric space of 
interest and present the spectral energy distributions and light
curves predicted by the model.
The visibility and temporal signature of the flare is 
analysed in Section \ref{sec:obsfeautures}.
Finally, we summarise 
 and discuss our findings in Section~\ref{sec:summary}.

\section{The emerging cocoon scenario}
\label{sec:model}

   \begin{figure}
   \centering
   \includegraphics[width=\hsize]{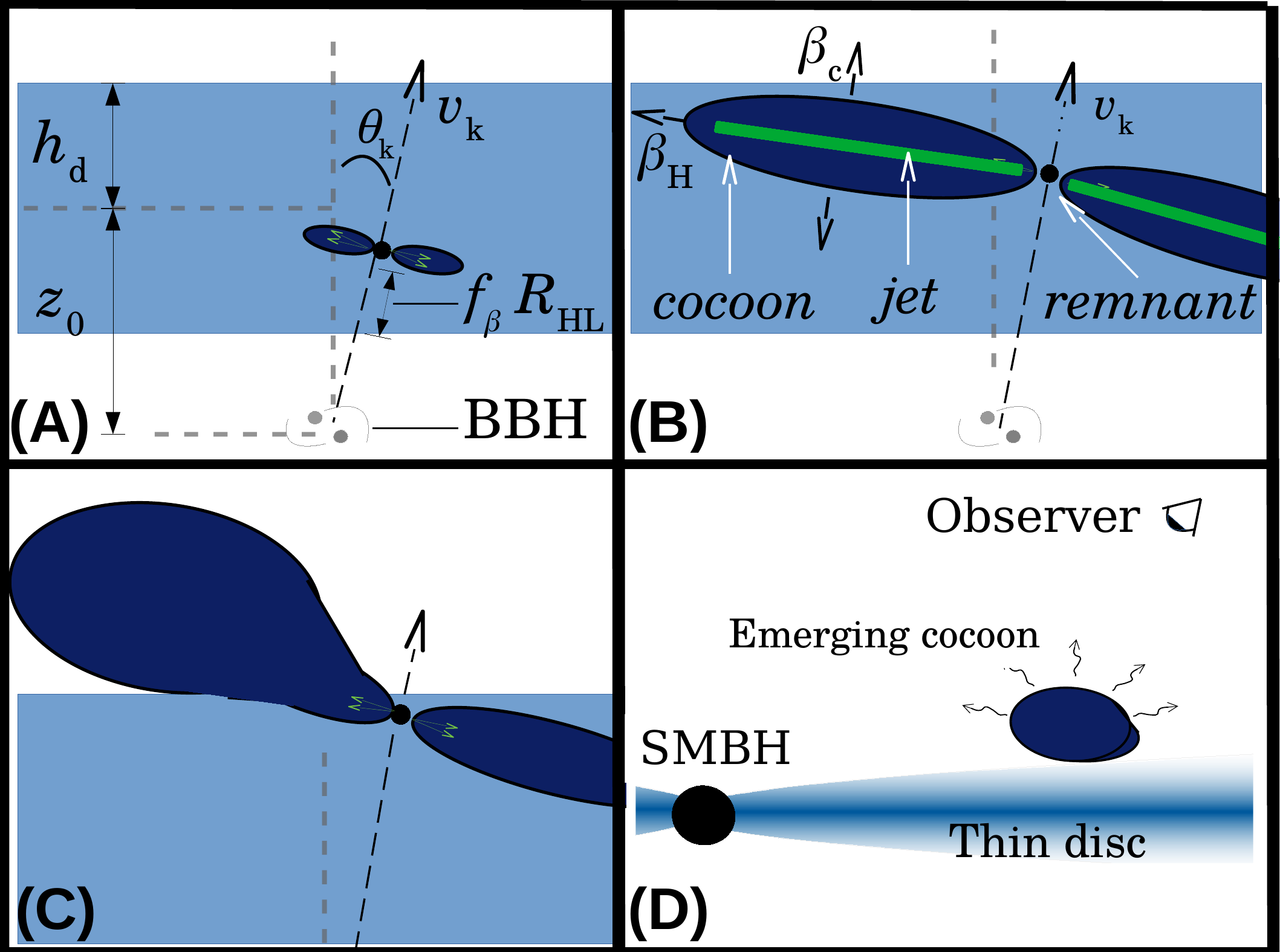}
      \caption{
Sketch of the scenario proposed in this work for the EM counterpart 
to a BBH GW event. 
The blue rectangular regions in panels (A) and (B) represent the 
region of the AGN thin disc of half-thickness $h_\mathrm{d}$ 
at a distance $a$ from the SMBH, nearby the location of the BBH merger.
(A): The BBH coalescence occurs at a vertical distance $z_0$
from the disc mid-plane, leaving a BH remnant with a gravitational 
kick of velocity $v_\mathrm{k}$.
The remnant BH launches efficient jets in a period $f_\beta R_\mathrm{HL}/v_\mathrm{k}$
after entering the disc.
(B): The jets inflate cocoons of shocked disc material (represented by the dark blue regions) 
that propagate within the disc.
(C): The material of one of the jet cocoons emerges and 
expands outside the disc on the observer side.
(D): The emerging cocoon let escape thermal photons producing an observable flare.
              }
         \label{fig:scenario}
   \end{figure}

Highly spinning BHs with kick velocities of $\sim$ [100-1000] Km s$^{-1}$
can result after coalescences of non-equal mass binaries 
(\citealt{2007PhRvL..98w1102C, 2007ApJ...659L...5C, 2015PhRvD..92b4022Z}).
In addition, spinning BHs accreting magnetised gas can launch jets with 
100\% efficiency, or higher, due to extraction of spin energy 
as described by the Blandford-Znajek mechanism 
(\citealt{1977MNRAS.179..433B, 2012MNRAS.423.3083M, 2022arXiv220111753K}).
Motivated by the aforementioned theoretical predictions,
here we analyse the situation of a spinning BH remnant of BBH coalescence 
that enters the dense region of an AGN thin disc and
launches relativistic jets while travelling within it.

We focus on BH remnants from mergers of second or higher generations, which can be viable explanations for high-mass remnants inferred by LVC detections, like GW170729 \citep{PhysRevX.9.031040} and GW190521 \citep{2020ApJ...900L..13A}.
Since the components of a second-generation merger were produced in previous coalescences, these components were likely born with post-merger kicks that set them outside of the disc for a significant fraction of their orbital period, as we estimate in Appendix~\ref{app:kzmax}. 
When such kicked remnants then form a new binary system, the orbit of the latter could also be misaligned with the disc plane.
This motivates us to explore the case of a second (or higher) generation merger occurring outside the AGN disc.
Thus, in the present analysis we consider a BBH coalescence 
taking place at a distance $a$ from the 
AGN SMBH, at a vertical distance from the disc mid-plane $z_0$, being
$h_\mathrm{d}(a)<z_0 < a$, 
with $h_\mathrm{d}(a)$ the disc semi-thickness at the radial distance $a$. A kicked remnant of mass $M_\bullet$ is produced at coalescence with instantaneous
kick velocity $v_\mathrm{k}$ forming an angle $\theta_\mathrm{k}$ with respect 
to the disc normal, as depicted in Figure~\ref{fig:scenario}A.
After the coalescence, 
the BH remnant meets the disc surface in the
(source frame) period
\begin{equation}
\label{Atent}
\Delta t_\mathrm{ent} = 
\frac{z_0 - h_\mathrm{d}}{v_\mathrm{k} \cos{\theta_\mathrm{k}}}.
\end{equation}
Once the remnant enters the disc, 
it undergoes Bondi-Hoyle-Lyttleton (BHL) accretion 
(\citealt{1944MNRAS.104..273B,2014ApJ...783...50L, 2015ApJS..219...30L}) of the disc material.
Such accretion drives relativistic jets which take to form 
a period of
\begin{equation}
\label{Atj}
\Delta t_\mathrm{j} = f_\beta R_\mathrm{HL}/v_\mathrm{k},
\end{equation}
after the BH enters the disc. 
In equation (\ref{Atj}),
\begin{equation}
R_\mathrm{HL}=2GM_\bullet/v_\mathrm{k}^2,  
\end{equation}
is the so-called BHL radius, and $f_\beta$ is a factor that depends
on the magnetisation of the external medium.
In this work we consider BH remnants launching jets with the fiducial 
efficiency of 200\%.
According to general relativistic magneto-hydrodynamical (GRMHD)
simulations of wind accretion performed by \cite{2023ApJ...950...31K},
jets with 200\% efficiency can be launched by BHs travelling within a gaseous
environment of plasma beta $\beta\sim 10$.
Such efficient jets take a period of  
$\sim 2 R_\mathrm{HL}/v_\mathrm{k}$ to form, as can be seen in 
Figures~3 and 4 of \cite{2023ApJ...950...31K}.
Thus, we consider the fixed value of $f_\beta = 2$
through this paper.

As the BH trajectory proceeds, the jets propagate within the disc 
inflating bipolar cocoons (\citealt{2011ApJ...740..100B})
of shocked disc material that eventually meets the edge of the disc (see Figure~\ref{fig:scenario}B).
Hence, we consider the jet cocoons as the channel
through which the 
BH remnant transports mass and energy outside the disc.
Here we refer to the cocoon material that emerges from the disc as the ``emerging cocoon''.

When the cocoon meets the disc boundary
(defined by the disc semi-height), its jet-like morphology
is no longer preserved 
due to the drastic drop of gas density beyond the disc semi-height $h_\mathrm{d}$\footnote{Gaussian profiles along the z direction are suitable disc solutions of 
the vertical disc structure}.
At this point, the cocoon material prefers to expand laterally
on the side where the disc density drops,
as depicted in Figure~\ref{fig:scenario}C.
Here we focus on emerging cocoons driven by jets propagating
quasi-parallel to the disc plane. 
We note that compared to jets propagating 
in a quasi-perpendicular direction,
quasi-parallel jets have the chance to sweep up more
disc material and their cocoons break out the disc edge
with a slower flow velocity along the disc's vertical direction 
(quasi-perpendicular jets would drive faster and more
jetted-like outflows).
Thus, we approximate the emerging cocoon
as a quasi-spherical, non-relativistic expanding flow.
When breaking out the disc, this outflow could produce a brief high-energy
EM transient due to  the acceleration of non-thermal
particles at the outflow expansion front. 
Here we are, however, focused on the thermal emission produced
by photons that diffuse within the emerging cocoon
and emanates from its outer surface.

The emerging cocoon is composed by swept-up
disc material collected while the BH travels within the disc.
At the same time, the jet that creates the cocoon takes the period given by the equation (\ref{Atj}) to form.
Therefore, the production of the emerging cocoon is constrained
by the thickness of the disc where the remnant plunges.
In the present analysis, we then define 
\begin{equation}
h_\mathrm{d}>f_\beta R_\mathrm{HL},
\label{hdcond}
\end{equation}
as a necessary condition for the production of emerging cocoons.

In the next subsection, we calculate the mass and energy transported 
by the cocoon outside the disc on the observer side and in 
Subsection~\ref{subsec:outflow-exp}, we assess the expansion of the 
emerging cocoon and its thermal emission.

\subsection{The jet cocoon propagation within the thin disc}
\label{subsec:jet-cocoon}

We consider the energy of the emerging cocoon as equivalent to  
the energy injected by the travelling BH through one of its jets and within the period when the BH undergoes BHL accretion of the disc material.
For analytic purposes, we assume uniform gas density within the disc
(with a thickness of semi-height $h_\mathrm{d}$) and 
drastically smaller density outside.
We then estimate the energy stored in the 
emerging cocoon as 
\begin{equation}
\label{E0}
E_0 = L_\mathrm{j}\Delta t_\mathrm{bh},
\end{equation}
where $L_\mathrm{j}$ 
is the BH jet power and $\Delta t_\mathrm{bh}$ is the time interval since the jet is formed
until the BH reaches the AGN disc boundary 
(see Figure~\ref{fig:scenario}).
The time interval of jet energy injection then is 
\begin{equation}
\label{Atbh}
\Delta t_\mathrm{bh} = \frac{2 h_d }{v_\mathrm{k}\cos\theta_\mathrm{k}} - \frac{f_\beta R_\mathrm{HL}}{v_k},
\end{equation}
being the first term in the RHS of equation (\ref{Atbh}) the time 
that the BH takes
to cross the disc of
thickness 
$2 h_\mathrm{d}$,
and the second term the time  needed for the jet to form 
(see equation~\ref{Atj}).

Motivated by the results of \cite{2022arXiv220111753K}, we take the 
power of one of the BH jets as
\begin{equation}
\label{Lj}
L_\mathrm{j} = \dot{M}_{\bullet} c^2,
\end{equation}
which corresponds to a total jet power with an efficiency of 200\% 
(the BH launches two jets), being 
\begin{equation}
\label{accrate}
\dot{M}_{\bullet} = f_\mathrm{acc} 4\pi\rho_\mathrm{d} \frac{(GM_\mathrm{\bullet})^2}{(c_\mathrm{s}^2 + v_\mathrm{k}^2)^{3/2}},
\end{equation}
the accretion rate onto the travelling BH.
The RHS of equation (\ref{accrate}) represents a fraction $f_\mathrm{acc}$ of the BHL 
accretion rate onto a massive particle travelling within a non-magnetised medium with sound speed
$c_\mathrm{s}$ and gas density $\rho_\mathrm{d}$.
Through this paper we adopt the fixed value of $f_\mathrm{acc}=0.1$, which is consistent
with a magnetised medium of  $\beta\sim 10$, according to 
\cite{2023ApJ...950...31K}.

We estimate the mass of the emerging cocoon as the mass of the material enclosed within 
the jet cocoon just before breaking out the disc.
The cocoon encloses a mixture of jet-ejected material
plus swept-up disc material.
We assume the jet-ejected gas to be much more diluted
compared to the disc material, and that the mass of the cocoon is always
dominated by the swept-up mass when the cocoon meets the 
disc edge. Then we estimate the mass of the cocoon as follows:
\begin{equation}
\label{Mc}
M_\mathrm{c} = \rho_d \pi r_c^2(\Delta t_c) z_\mathrm{H}(\Delta t_c),
\end{equation}
where we approximate the volume of the cocoon as a cylinder of height $z_\mathrm{H}$ and radius $r_\mathrm{c}$, being
$\Delta t_c$ the time interval since the jet is formed (see equation ~\ref{Atj})
until the cocoon reaches the AGN disc edge (see Figure~\ref{fig:scenario}B).
To obtain the height and radius of the cocoon, we employ the formalism of \cite{2011ApJ...740..100B},
for the case of a relativistic jet propagating through a uniform medium. Then,
we note that the period $\Delta t_\textrm{c}$ is related to the disc half-thickness and the
remnant's velocity as:
\begin{align}
\nonumber
& 2 h_\mathrm{d} - f_\beta R_\mathrm{HL} \cos\theta_\mathrm{k}   = \\
\label{eqtc}
& \frac{r_\mathrm{c}(\Delta t_\mathrm{c})}{\cos\theta_\mathrm{k} } + \sin\theta_\mathrm{k}
[z_\mathrm{H}(\Delta t_\mathrm{c}) - \tan\theta_\mathrm{k} r_\mathrm{c}(\Delta t_\mathrm{c}) ] +
\Delta t_\mathrm{c} v_\mathrm{k} \cos\theta_\mathrm{k},
\end{align}
where $z_\mathrm{H}$ and $r_\mathrm{c}$ are the height and 
the cylindrical radius of the cocoon, parameterised as:
\begin{equation}
\label{zHrc}
z_\mathrm{H} = c \int_0^{\Delta t_c} dt \beta_\mathrm{H}(t), \,\,\,\,r_\mathrm{c} = c \int_0^{\Delta t_c} dt \beta_\mathrm{c}(t).
\end{equation}

We adopt the  solutions derived by \cite{2011ApJ...740..100B}
to obtain the speeds of growth $\beta_\mathrm{H}$ and $\beta_\mathrm{c}$
(in units of the speed of light) for the height and radius of the cocoon, respectively:
\begin{align}
\label{betaH}
&\beta_\mathrm{H}(t) = 
\left[ 
1 + \tilde{L}(t)^{-1/2}
\right]^{-1},\\
\label{betac}
&\beta_\mathrm{c}(t) = \frac{\theta_0}{2}\tilde{L}(t)^{1/2},\\
\label{tilL}
&\tilde{L} = A\left(\frac{L_\mathrm{j}}{\rho_\mathrm{d} \theta_0^4 c^5}\right)^{2/5} t^{-4/5},
\end{align}
with $A\sim 0.7$, and $\theta_0$ the jet opening angle at the base. 
To obtain $z_\mathrm{H}$, $r_\mathrm{c}$ and hence the mass within the cocoon
(equation \ref{Mc}), one can first solve equation (\ref{eqtc})
to obtain $\Delta t_c$ and then evaluate  equations (\ref{zHrc}). 
We adopt the standard thin disc model of
\cite{1973A&A....24..337S} (hereafter SS), 
to parameterise the
properties of the disc
(such as the semi-height $h_\mathrm{d}$, density $\rho_\mathrm{d}$, and
speed of sound $c_\mathrm{s}$)
as a function of 
the distance to the SMBH $a$,
and its mass $M_\mathrm{s}$ and accretion rate $\dot{M}_\mathrm{s}$.

The cocoon solution given by equations (\ref{betaH})-(\ref{betac})
corresponds to the ``collimated jet'' regime
\citep{2011ApJ...740..100B}.
This is a suitable description for the cocoon evolution as long as $\tilde{L}<1$.
If on the other hand $\tilde{L}>\theta_0^{-4/3}$, one should 
adopt the mathematical solutions corresponding to the
``uncollimated jet''. 
Combining equations (\ref{Lj}), (\ref{accrate}), 
and (\ref{tilL}) we estimate
\begin{equation}
\label{tildeLest}
\tilde{L} \sim 0.114
\left(
\frac{\theta_0}{15^{\circ}}
\right)^{-8/5}
\left(
\frac{M_\bullet}{200\,\mathrm{M}_\odot}
\right)^{4/5}
\left(
\frac{v_\mathrm{k}}{200\,\mathrm{Km}\,\mathrm{s}^{-1} }
\right)^{-6/5}
\left(
\frac{\Delta t_\mathrm{c}}{0.1\,\mathrm{day}}
\right)^{-4/5}.
\end{equation}
In this work we explore EM counterparts 
from BH remnants of
$M_\bullet \lesssim 200$ M$_\odot$ and
$v_\mathrm{k} \gtrsim 200\,\mathrm{Km}\,\mathrm{s}^{-1}$.
In addition, we find that solutions to
equation (\ref{eqtc}) 
give in general $\Delta t_\mathrm{c}\gg0.1\,\mathrm{day}$.
Thus, the parameter configurations considered here lead in general to $\tilde{L}\ll 1$
and we then consider the collimated jet solution only throughout this work.

\subsection{The expansion of the emerging cocoon}
\label{subsec:outflow-exp}

The outflow of disc material that emerges from the disc
is a mixture of matter and photons that produce an EM flare after expanding
enough to let escape the thermal photons.
For calculating such emission, here we model the evolution of the 
emerging cocoon 
as equivalent to an expanding sphere of mass $M_0$,
of initial uniform density $\rho_0$, and total energy $E_0$, similarly to a supernova remnant.

We take the mass and initial volume
of the expanding sphere
as the mass and volume of the jet cocoon just before
breaking out the disc
(see equation \ref{Mc} and related text
in the previous section), and hence we take the
initial density and radius as 
$\rho_0 = M_c/V_0$ and $r_0 = [3V_0/(4\pi)]^{1/3}$, respectively.
We assume the total energy $E_0$ (given by equation \ref{E0}) 
to split into kinetic and thermal energies during the outflow expansion:
\begin{equation}
\label{Ek}
E_\mathrm{k} = \alpha_\mathrm{k} E_0, 
\end{equation}
\begin{equation}
\label{Eth}
E_\mathrm{th} = (1-\alpha_\mathrm{k}) E_0,
\end{equation}
respectively and we assume energy equipartition setting $\alpha_\mathrm{k}=1/2$. 
The emerging cocoon is radiation pressure dominated when breaking out the disc. 
Thus the initial pressure $P_0$, and temperature $T_0$ can be related as
\begin{equation}
P_0 = a_\mathrm{r} T_0^4/3,
\end{equation}
being $a_\mathrm{r}$ the radiation constant.
Simultaneously, the initial pressure and volume 
can be related to the thermal energy (equation \ref{Eth} ) as
\begin{equation}
P_0 V_0/(\gamma_\mathrm{a} - 1) = E_\mathrm{th},
\end{equation}
being $\gamma_\mathrm{a}=4/3$ the adiabatic index appropriate
for a radiation pressure dominated gas.
Since the density of the environment outside the disc is negligible
compared to the outflow density, we consider a free expansion
for the outflow outer radius $R = R_0 + u_0 t'$, with the constant radial
velocity 
\begin{equation}
\label{u0}
u_\mathrm{0} = 
\sqrt{2 E_\mathrm{k}/M_0}.
\end{equation}

We calculate the emission of the emerging cocoon as that of a supernova (SN) remnant in its free expansion phase following
\citep{1980ApJ...237..541A,1996snih.book.....A,2012ApJ...746..121C}.
In this approach, the emission of the spherical expanding plasma
is produced by photons arriving by diffusion at the outflow surface.
The peak of bolometric luminosity occurs when the 
diffusion timescale equals the dynamic timescale. 
Following the approach of 
\citep{1996snih.book.....A,2012ApJ...746..121C}, 
the diffusion time is
\begin{equation}
t_\mathrm{d} = \frac{3R^2\rho \kappa}{\pi^2c} = \frac{\kappa M_0}{b c R},
\end{equation}
where $\kappa$ is the gas opacity taken as constant, and
$b = 4\pi^3/9$.
Considering the outflow dynamic time as $t_\mathrm{h} = R/u_0$,
the condition $t_\mathrm{d} = t_\mathrm{h}$ occurs at 
\begin{equation}
\label{tmax}
t_\mathrm{max} = \sqrt{\frac{\kappa M_0 }{b c u_0}}.
\end{equation}
Henceforth, the bolometric luminosity evolves as \citep{1996snih.book.....A,2012ApJ...746..121C}:
\begin{equation}
\label{Lt}
L(t') = \frac{4\pi a_r b c}{3}\frac{T_0^4 R_0^4}{\kappa M_0}
\exp \left\{
-\left[
\frac{(t' - t_\mathrm{max})^2}{t_g^2} + \frac{2R_0(t' - t_\mathrm{max})}{u_0t_g^2}
\right]
\right\},
\end{equation}
with $t_g = \sqrt{2 t_\mathrm{d,0}t_\mathrm{h,0}}$, 
$t_\mathrm{d,0} = \kappa M  / (bcR_0)$, and 
$t_\mathrm{h,0} = R_0 / u_0$.
We take $\kappa$ as the electron opacity
$\kappa_\mathrm{T} = \sigma_\mathrm{T}/(\mu_\mathrm{e} m_\mathrm{u})$,
being $\sigma_\mathrm{T}$ the electron
scattering cross-section, $m_\mathrm{u}$ the atomic mass constant,
$\mu_\mathrm{e} = 2/(1 + X)$, and we use $X = 0.85$ as the hydrogen 
mass fraction.
We then calculate the spectrum emitted by the emerging cocoon as 
black-body radiation of effective temperature
\begin{equation}
T_\mathrm{eff}(t') = 
\left[
\frac{L(t')}{4\pi R(t')^2 \sigma_\mathrm{SB}}
\right]^{1/4},
\end{equation}
being $\sigma_\mathrm{SB}$ the Stefan-Boltzmann constant.
Thus, the flux density of radiation at the time $t$ and frequency $\nu$
in the observer frame (at Earth) is 
\begin{align}
\label{Snu}
&\nu F_\nu(t) = \nu'
\pi \left[ \frac{R(t')}{D_\mathrm{L}^2}\right]^2 
B_{\nu'}\left[T_\mathrm{eff}(t')\right], \\
&t'= \frac{t}{1+z},\\ &\nu'= (1+z)\nu,
\end{align}
being $z$ and $D_\mathrm{L}$
the source red-shift and luminosity distance, respectively,
and $B_{\nu'}$ is the black-body spectral radiance of temperature
$T_\mathrm{eff}$.

\section{AGN+emerging cocoon emission profiles}
\label{sec:profiles}

To investigate the wavelengths at which the outflow
flare discussed in Section ~\ref{sec:model} can be observed,
we compare its spectrum with the emission of the 
hosting AGN.
To obtain a particular emission profile, we first specify 
assumed values for
$z$ (redshift of the source), 
$D_\mathrm{L}$ (luminosity distance),
$M_\mathrm{s}$ (SMBH mass),
$\dot{M}_\mathrm{s}$ (SMBH accretion rate), 
$\alpha$, (disc viscosity parameter), and 
$a$ (distance from the SBMH where the merger occurs).
Based on these parameters, we derive through
the standard disc model of SS,
the properties of the local  environment where the
remnant BH interacts, namely
$\rho_\mathrm{d}$ (gas density),
$T_\mathrm{d}$ (temperature), and
$h_\mathrm{d}$ (disc half thickness).
Then, specifying 
the parameters
of the remnant, namely 
$M_\bullet$ (BH mass)
$v_\mathrm{k}$ (kick velocity), 
$\theta_\mathrm{k}$ (kick angle relative to the disc normal), and
$\theta_0$ (jet opening angle), we 
obtain the
thermal emission from the 
emerging cocoon
as described in Section~\ref{sec:model}.

Given the chosen values of $M_\mathrm{s}$ and $\dot{M}_\mathrm{s}$, we assess 
the emission of the hosting AGN as
$
\nu L_{\nu, \mathrm{bg}} = 
(f_\mathrm{n}/l_\mathrm{ref}) \nu L_{\nu, \mathrm{ref}},
$
where
$\nu L_{\nu,\mathrm{ref}}$ is the average 
AGN spectrum profile given by the blue or cyan points in Figure 7 
\citep{2008ARA&A..46..475H}, $l_\mathrm{ref}$ is the reference
luminosity of this spectrum at $\lambda_\mathrm{ref} = 4400 $ \AA,
and $f_\mathrm{n}$ is a normalisation factor
that depends on the mass and accretion rate onto the SMBH and that we define
following \cite{2023ApJ...950...13T} as:
\begin{equation}
f_\mathrm{n} = 10^{44}\mathrm{erg}\,\mathrm{s}^{-1}
\left(
\frac{M_\mathrm{s}}{10^8 \mathrm{M}_\odot}
\right) 
\left(
\frac{\dot{M}_\mathrm{s} c^2}{L_\mathrm{Edd}(M_\mathrm{s})}\right)
\left(
\frac{10}{f_\mathrm{c}}
\right).
\label{fn}
\end{equation}
In this normalisation, we use the value of $f_\mathrm{c}=3$, leading to a background AGN spectrum
consistent with the calculated disc emission.
Given the values of 
$M_\mathrm{s}$, $\dot{M}_\mathrm{s}$, and $\alpha$, 
we calculate the disc emission 
by integrating the black body 
radiation of the disc surface (see e. g., \citealt{2002apa..book.....F}),
from the innermost stable circular orbit of a non-spining SMBH, $6 R_\mathrm{g}$, to $10^{4} R_\mathrm{g}$.

Motivated by the  location of the
AGN J124942.3+344929, claimed as the
host of the first EM counterpart to a BBH event \citep{2020PhRvL.124y1102G},
throughout this paper we use the fiducial values  of 
$z = 0.438$ and
$D_\mathrm{L}= 1734.166$ Mpc,
for the redshift and luminosity distance of the 
source, 
respectively\footnote{
Given the redshift $z$, we
use the package \texttt{astropy} to
estimate the luminosity distance $D_\mathrm{L}$ 
through a $\Lambda$CDM cosmology
together with Planck 2018 results 
\citep{2020A&A...641A...6P}.}.
The location where the BH remnant might interact with the
disc is so far not well constrained, thus we explore multiple locations within the range of
$a=[1000-8000] R_\mathrm{g}$ ($R_\mathrm{g} = GM_\mathrm{s}/c^2$),
which are of the order of the migration trap location
discussed in \cite{2016ApJ...819L..17B}. 

In order to assess the disc properties at the above radii,
we employ the disc model of SS and consider accretion rates within [0.01, 0.1] $\dot{M}_\mathrm{Edd}$ with the Eddington accretion rate defined as $\dot{M}_\mathrm{Edd}=1.39\times 10^{18} [M/M_\odot]$ g s$^{-1}$.
Since thin accretion discs can be gravitationally unstable at parsec scales, we evaluated the radius within which the disk is stable against self-gravity varying the SMBH mass between $10^6$ and $10^{10} M_\odot$ \citep{1990AA...229..292C}. We find that SS discs are stable for SMBHs with $M_s \lesssim 10^8$ M$_\odot$ at the radii and accretion rates considered here. Thus, in this paper, we explore EM counterparts restricted to SMBHs ranging [$5\times 10^6 - 5\times 10^7$] M$_\odot$. This choice for the SMBH mass range is also be motivated by the number density of SMBHs at the local Universe, which is higher for masses $\sim[10^6 -10^7]$ M$_\odot$ than for masses $\gtrsim 10^8$ M$_\odot$ \citep{2014ApJ...786..104U}.

We assume that recoils from previous mergers slightly perturbed the
alignment with the disc of the binary components discussed here,
and thus, we consider the merger to occur outside the disc.
The vertical location of the merger $z_0$
is then restricted to the maximum displacement from the disc 
mid-plane that the binary components attained due to the 
kicks of their previous coalescences.
In AGNs with SMBHs of masses $\gtrsim 5\times 10^{6}$ M$_\odot$, remnants of
mergers occurring within the disc at
$\sim$[1000 - 8000] $R_\mathrm{g}$ from the SMBH and with kick velocities $\lesssim 1000$ Km s$^{-1}$, are retained by the AGN gravitational potential.
Such kicked BHs can reach a maximum vertical displacement 
from the disc mid-plane ranging about $\sim$ [5-100] times of the 
disc semi-height $h_\mathrm{d}$, as we estimate
in Appendix~\ref{app:kzmax}.
This maximum $z-$displacement can be reduced due to dynamical friction acting every time the kicked BH crosses the thin disc.
Detailed calculation of the remnant's orbit considering the interacting with the AGN disc
is beyond the scope of the present work.
Thus, we consider the initial position of the merger considered here
(a second or higher order merger generation) as a free parameter.
To illustrate the solutions derived from the present multi-messenger
scenario, we consider the values of $z_0/h_\mathrm{d} =5$ and 20, which
are consistent with the maximum $z-$displacements derived
in Appendix~\ref{app:kzmax}.

Assuming that the jet propagates quasi-perpendicular to the 
BH trajectory,
we choose the small angle of $\theta_\mathrm{k} = 8^{\circ}$, since we are focusing
on jets propagating quasi-parallel to 
the disc plane (see Section~\ref{sec:model}).
For the jet opening angle, we use the fixed value of   
$\theta_0=15^{\circ}$ 
in all calculations in this work.
We explore emission profiles corresponding to 
BHs with masses  and kick velocities in the ranges of  [50-200] M$_\odot$ 
and [100-1000] Km s$^{-1}$, respectively,  motivated by potential measures of BHs 
by the LVK experiment \citep{2020PhRvL.125j1102A}
as well as by numerical models of non-symmetrical BBHs mergers
\citep{2015PhRvD..92b4022Z}.

In the following subsection, we discuss the time delay 
at the Earth frame for the appearance of the EM counterpart 
and in Subsections \ref{ssec:seds} 
and \ref{ssec:lcs}, we present spectral energy distributions
(SEDs) and light curves (LCs),
respectively, of the flare emission.

\subsection{The flare starting time}

\begin{figure}
   \centering
   \includegraphics[width=\hsize]{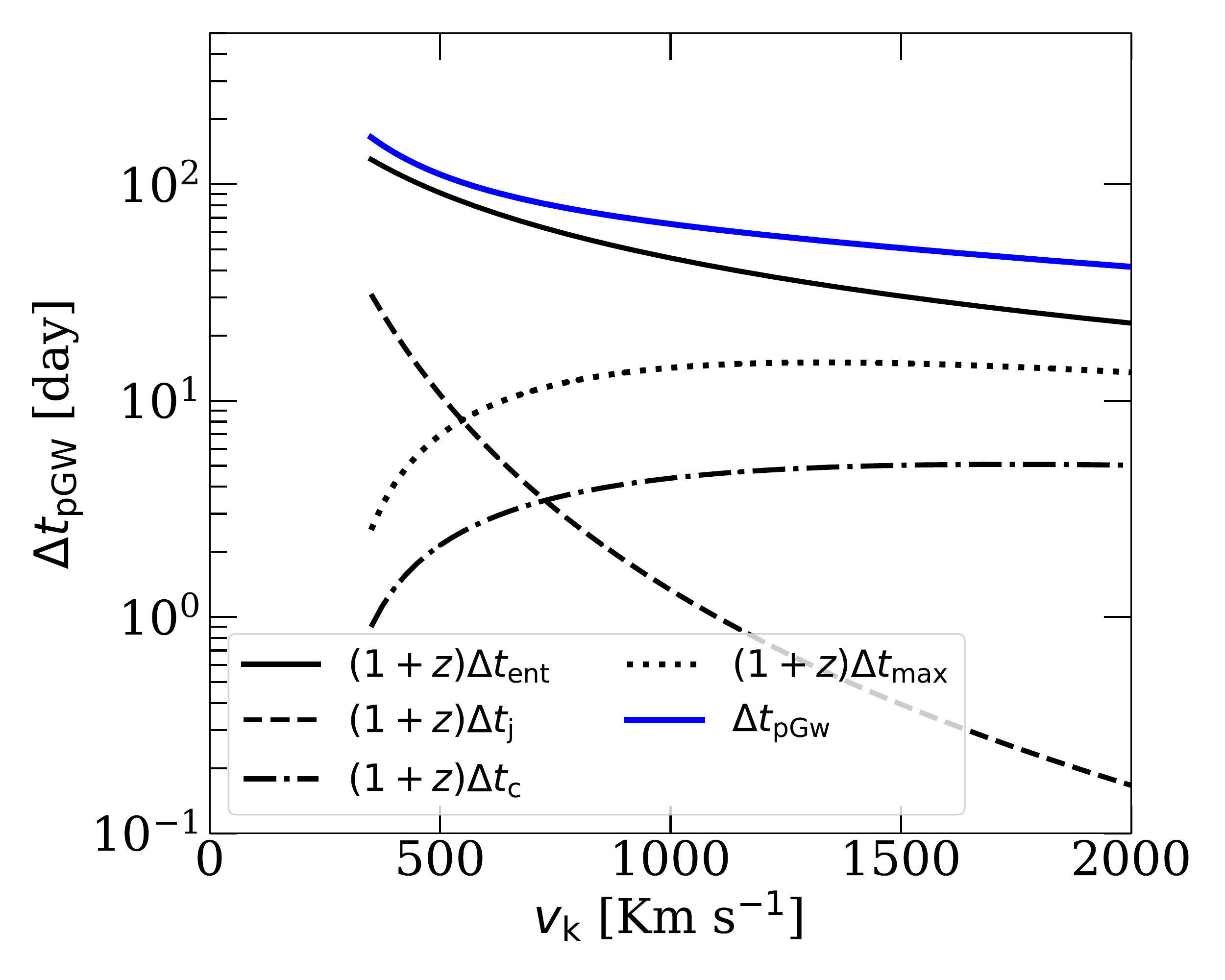}
      \caption{
Time delay $\Delta t_\mathrm{pGW}$ (in the observer frame) as a function of the 
remnant kick velocity
for the appearance of the emerging cocoon flare after the GW event.
This time interval is a sequence of
different periods of the present emission model (see the text), namely
the periods in which
(i) the BH remnant enters the disc boundary, $\Delta t_\mathrm{ent}$,
(ii) the jets form within the disc, $\Delta t_\mathrm{j}$, 
(iii) the jet cocoon reaches the disc boundary, $\Delta t_\mathrm{c}$, and
(iv) the emerging cocoon produces the maximum bolometric luminosity $\Delta t_\mathrm{max}$ (see the text).
The blue curve plots $\Delta t_\mathrm{pGW}$ as the sum of the aforementioned
period components.
}
         \label{fig:Atpgw_comps}
   \end{figure}

In the multi-messenger scenario discussed here, 
a BBH coalescence produces a GW signal followed by an EM flare
starting after a time delay
 $\Delta t_\mathrm{pGW}$.
This time delay is computed as:
\begin{equation}
\label{Atpgw}
\Delta t_\mathrm{pGW} = (1+z)  
(
\Delta t_\mathrm{ent} +
\Delta t_\mathrm{j}   +
\Delta t_\mathrm{c}   +
\Delta t_\mathrm{max}
),
\end{equation}
which comprises  the
temporal periods in which 
(i) the BH remnant enters the disc 
$\Delta t_\mathrm{ent}$ (equation \ref{Atent}), 
(ii) the jet forms within the disc 
$\Delta t_\mathrm{j}$ (equation \ref{Atj}),
(iii) the jet cocoon reaches the disc boundary 
$\Delta t_\mathrm{c}$ (equation \ref{eqtc}), and
(iv) the emerging cocoon produces its maximum bolometric luminosity
$\Delta t_\mathrm{max}$ (see equation \ref{tmax}).
In equation (\ref{Atpgw}),
the  factor $(1+z)$ accounts for the time dilation due to the red shift of the source. We illustrate in Figure~\ref{fig:Atpgw_comps} 
the delay $\Delta t_\mathrm{pGW}$ and its
components as a function of the kick velocity.
This example corresponds to a remnant of 
150 M$_\odot$, from a coalescence at
$z_0 = 5 h_\mathrm{d}$ and $a = 5000 R_\mathrm{g}$
from an SMBH of $3\times 10^{7}$ M$_\odot$ accreating at 0.03 $\dot{M}_\mathrm{Edd}$.

We note that in the limit of $z_0\gg h_\mathrm{d}$
the $\Delta t_\mathrm{ent}$ component dominates in
the RHS of
equation (\ref{Atpgw}). 
In this limit, we can approximate
$\Delta t_\mathrm{ent} 
\approx (z_0/v_\mathrm{k})(1-h_\mathrm{d}/z_0)/(1-\theta_\mathrm{k}^2)
\approx z_0/v_\mathrm{k}$.
The other three components represent a subdominant
contribution to the delay time $\Delta t_\mathrm{pGW}$ 
and we simply approximate them as
the time that the remnant spends within 
the disc 
$\approx 2h_\mathrm{d} / v_\mathrm{k}$.
Thus in the $z_0\gg h_\mathrm{d}$ limit,
we approximate $\Delta t_\mathrm{pGW}$ as:
\begin{equation}
\label{Atpgw_app}
\Delta t_\mathrm{pGW,0} = (1+z)  
\frac{z_0}{v_\mathrm{k}}
\left(
1+\frac{2h_\mathrm{d}}{z_0}
\right).
\end{equation}

\begin{figure}
   \centering
   \includegraphics[width=\hsize]{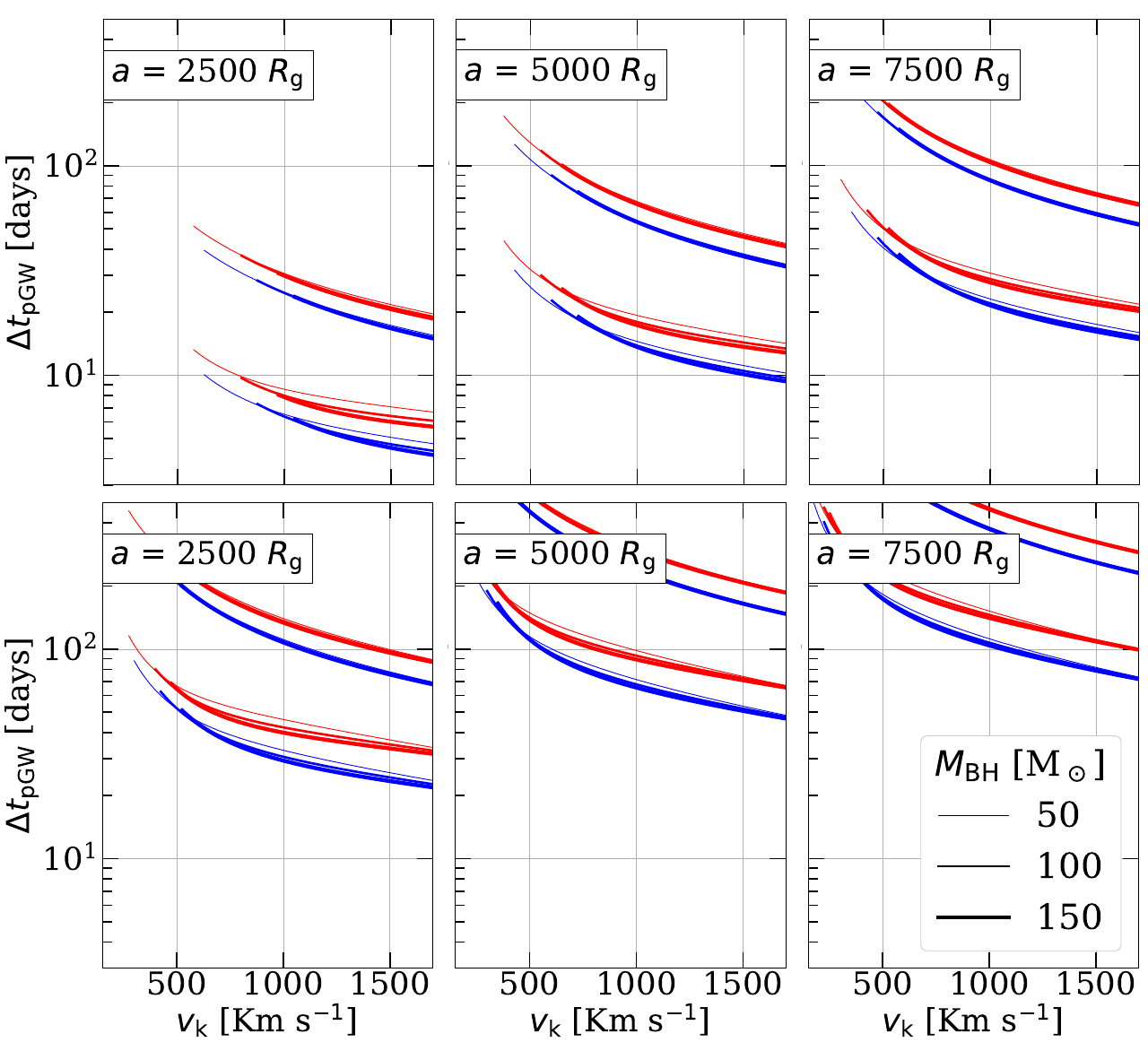}
      \caption{
Time delay for the appearance of the flare after the BBH coalescence (measured at Earth) derived
as a function of 
the kick velocity $v_\mathrm{k}$ of the merger remnant.
In each panel, the curves are calculated using a fixed distance $a$
as indicated.
Upper and lower panels are obtained assuming an SMBH 
of $6\times 10^{6}$ and $3\times 10^{7}$ M$_\odot$, respectively.
Blue and red curves correspond to the accretion rates of 
0.03 and 0.1, in Eddington units, respectively.
The sets of upper and lower curves in each panel are obtained assuming
$z_0 =$ 20 and 5$h_\mathrm{d}$, respectively. 
}
         \label{fig:Atpgw}
   \end{figure}

In Figure~\ref{fig:Atpgw}, we display the delay  $\Delta t_\mathrm{pGW}$ (equation~\ref{Atpgw}) as a function of the kick velocity,
for different values of the mass of the 
remnant
(indicated by the curve thickness), and different
SMBH masses  and accretion rates.
Upper and lower panels are obtained assuming 
$M_\mathrm{s}= 6\times10^6$ and $3\times10^7$ M$_\odot$,
respectively, whereas blue and red curves
correspond to 0.03 and 0.1 $\dot{M}_\mathrm{Edd}$, respectively.
In each panel, the upper and lower sets of curves correspond to
$z_0=20$ and 5$h_\mathrm{d}$, respectively.
We see that remnants with
kick velocities of $\sim$ [200-1500] 
Km s$^{-1}$ travelling 
through discs around SMBHs of
$5\times 10^{6-7}$ 
M$_\odot$ drives flares starting 
$\sim$[3-300] days after the GW signal, accordingly
We note that the delay $\Delta t_\mathrm{pGW}$ increases for larger 
mass and
accretion rate onto the SMBH.
The delay $t_\mathrm{pGW}$ is insensitive to the mass of the BH remnant
when $z_0$ is relatively large, as expected 
(see equation \ref{Atpgw_app}).

\subsection{Spectral energy distributions}
\label{ssec:seds}

   \begin{figure*}
   \centering
   \includegraphics[width=\hsize]{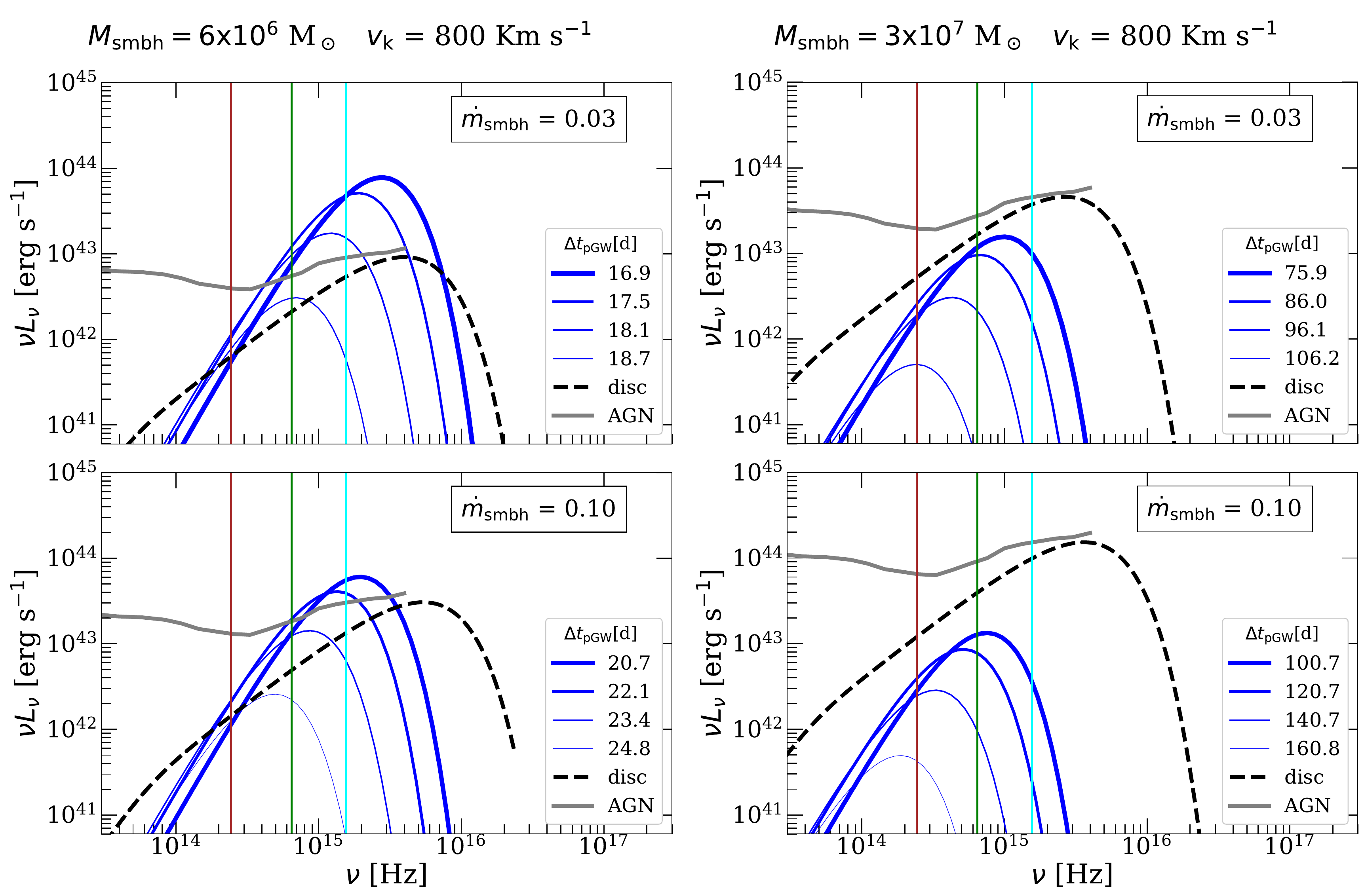}
      \caption{
SEDs of the emerging cocoon emission at different times after the assumed GW event (blue curves).
The curves were obtained through the model described in Section~\ref{sec:model} considering a
merger remnant 
of $100$ M$_\odot$ with a kick
velocity of $800$ Km s$^{-1}$
interacting with a thin disc 
at a distance of $a = 5000 R_\mathrm{g}$ from the central SMBH. 
Left and right panels assume an SMBH of $6\times10^{6}$ and $3\times10^{7}$ M$_\odot$, respectively, 
whereas the upper and lower panels consider
accretion rates of 0.03 and 0.1 (in Eddington units), respectively.
In each panel, the dashed black curve plots the emission of the associated thin disc.
The grey curves plot the background AGN emission 
adapted from \citealt{2008ARA&A..46..475H} (see the text).
The source is assumed to be located at a redshift of $z= 0.438$.
For reference, vertical lines indicate examples of frequencies at
NIR (J band, 1235 nm, brown), optical (g-band, 464 nm, green), and extreme ultraviolet (UVw2 band, 193 nm, cyan).} 
         \label{fig:seds}
   \end{figure*}

Figure~\ref{fig:seds}
displays spectral energy distributions (SEDs) corresponding to the differential luminosity of the emerging cocoon, the AGN disc,
and the AGN background emission
as measured at the Earth.
Blue curves are the outflow SEDs
at different time steps 
after the GW event, where the first SED curve
is calculated at the time when
the photon diffusion and the 
dynamical time of the emerging cocoon coincide
(see equation \ref{tmax}). 
The black dashed curves plot the stationary disc spectra,
and grey curves plot the AGN emission.
All curves in this figure were obtained assuming
a remnant BH of 100 M$_\odot$ with
kick velocity of 800 Km s$^{-1}$, interacting
at $a = 5000 R_\mathrm{g}$ and 
$z_0 = 5h_\mathrm{d}$ with the disc.
Panels on the left and right  correspond to an SMBH of
$6\times10^{6}$
and $3\times10^{7}$ M$_\odot$, respectively, whereas upper and lower panels
were obtained assuming accretion rates of 0.03 and 0.1 $\dot{M}_\mathrm{Edd}$
(onto the SMBH), respectively.

The examples of SED curves of Figure~\ref{fig:seds} illustrate some general
features of the EM counterpart flare described in Section~\ref{sec:model}. 
We note that for the relatively low accretion rate of 0.03 $\dot{M}_\mathrm{Edd}$ 
the flare can outshine or produce emission comparable to
the AGN  at NIR, optical and EUV wavelengths.
For the larger AGN accretion rate of $0.1\dot{M}_\mathrm{Edd}$, 
the flare emission peaks at lower frequencies.
The examples of SED profiles in Figure~\ref{fig:seds} 
then hints that the EM counterpart studied here could  
be well observed in EUV and optical, and perhaps in NIR.
In the following subsection, we describe the starting times and duration of such outflow flares by exploring  their LC profiles 
at NIR, optical, and EUV wavelengths for different masses and kick velocities of the BH remnant.

\subsection{Light-curves}
\label{ssec:lcs}

\begin{figure*}
   \centering
   \includegraphics[width=\hsize]{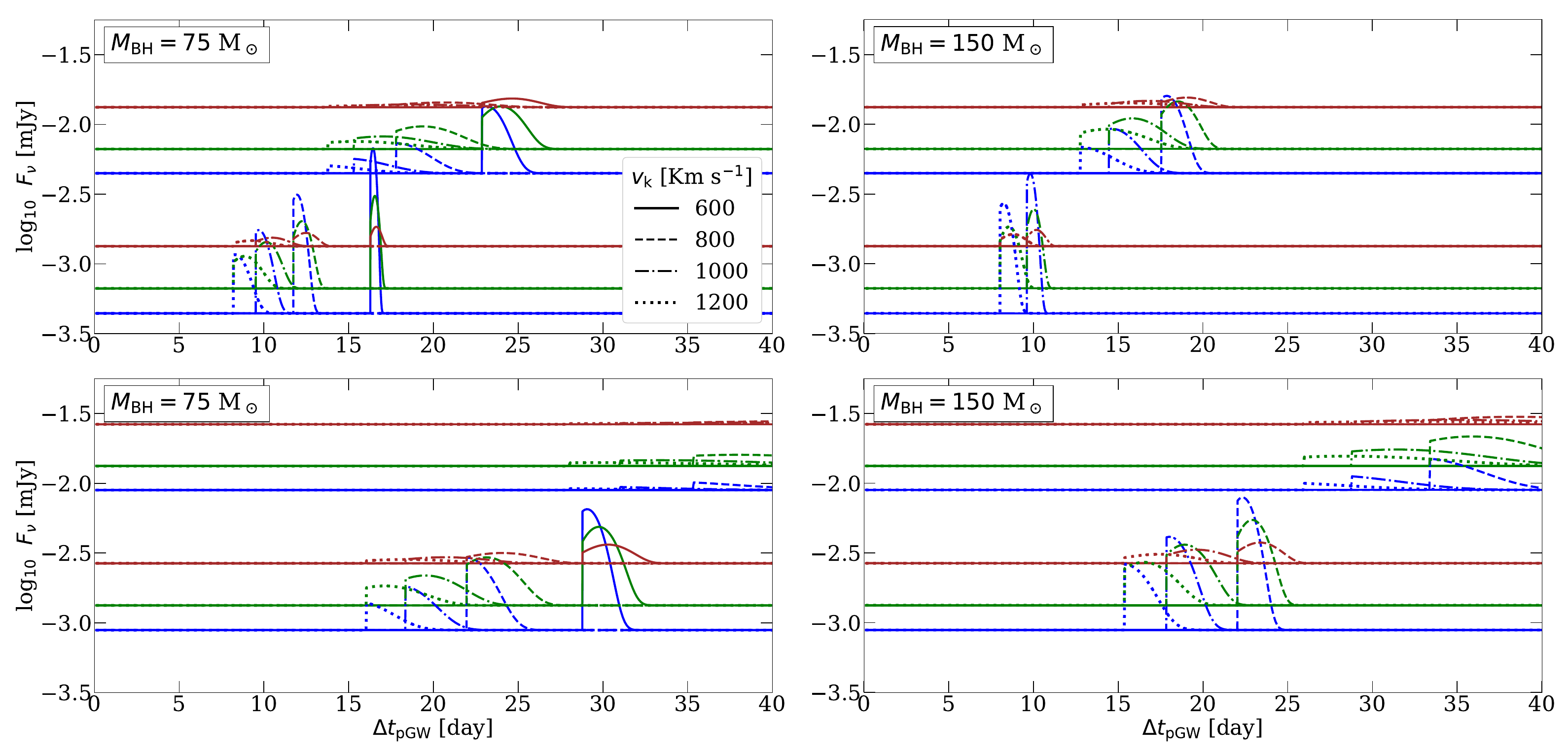}
      \caption{
LC profiles obtained from the 
emerging cocoon model described in Section~\ref{sec:model}.
The curves are flux densities at selected emission wavelengths, namely
NIR(J band, 1235 nm, brown), 
optical (g-band, 464 nm, green), and 
extreme ultraviolet (UVw2 band, 193 nm, darkcyan).
The LCs were obtained assuming an SMBH of $5\times 10^{6}$ M$_\odot$ (upper) and
$10^{7}$ M$_\odot$(lower).
As indicated, 
the left and right panels assume BH remnants 
of different masses and
different curve styles correspond to solutions with different BH kick velocities.
The upper and lower curves in each panel correspond to accretion rates of 0.1 and 0.01 
(in Eddington units), respectively, of the AGN central engine.
The AGN is assumed  located at a redshift of $z= 0.438$.
The timescale on the $x-$axis indicates the time elapsed,
in the observer frame,
after the assumed GW event.
              }
         \label{fig:NIR-EUV_LCsvks}
   \end{figure*}

We derive light-curve (LC) profiles
as a superposition of AGN plus flare emissions observed at Earth. 
For simplicity, we consider the AGN background emission
(grey curves in Figure~\ref{fig:seds}) as stationary, and 
we do not model the rise of the flare.
Instead, we add the flare to the background emission
at the time when the emerging cocoon emits its maximum
bolometric luminosity (see Subsection~\ref{subsec:outflow-exp}).
The obtained LC profiles are shown in Figure~\ref{fig:NIR-EUV_LCsvks} 
as flux per unit frequency where we illustrate examples of LCs at 
NIR, optical, and EUV frequencies.
The assumed GW event occurs at $t_\mathrm{pGW}=0$. 
The curves constantly start in time (stationary AGN emission) 
eventually displaying a flux jump followed by a temporal
evolution that converges to the initial background emission.
The left and right panels in Figure~\ref{fig:NIR-EUV_LCsvks} 
are obtained assuming a remnant 
BH of 75 and 150 M$_\odot$, respectively, whereas plots
in the upper and lower panels assume an SMBH of $5\times10^{6}$
and $10^{7}$ M$_\odot$, respectively. 
In each panel, LCs of fluxes at NIR, optical, and EUV frequencies  are plotted in
brown, green and blue colours, respectively, being the upper curve triads
associated to the SMBH accretion rate of  0.1 $M_\mathrm{Edd}$, 
and the lower triads to the 0.01 $M_\mathrm{Edd}$ rate.
Different line styles are associated
to different remnant kick velocities, as indicated.

The kick velocity has noticeable effects on the
flare starting time, amplitude, and duration.
For progressively larger values of $v_\mathrm{k}$ in the range of [600 - 1200]
Km s$^{-1}$, we note that in general: 
(i) the flare starting time decreases,
(ii) the flare amplitude decreases, 
(iii) the decaying flare period increases.
We also note that in AGNs with larger accretion rates onto the SMBH,
the emerging cocoon produces flares with larger decaying times. 
The mass of the BH remnant has the opposite effect: BHs with larger masses
produce flares with shorter decay times.


\section{Observational features}
\label{sec:obsfeautures}

AGNs typically exhibit continuum variability on timescales from weeks to years, currently of unknown origin. The emission variations of non-blazar AGNs are observed on the order of 10\% of their base emission in optical-UV wavelengths \citep{2004ApJ...601..692V, 2022ApJ...936..132Y}. This AGN intrinsic variability can challenge the identification of the flares derived in this work as GW counterparts.
In addition, different astrophysical processes can also produce flaring and enhancements in the AGN emission, such as variations in the accretion onto the SMBH \citep{2018MNRAS.480.4468R}, magnetic reconnection in the vicinity of the SMBH \citep{2010A&A...518A...5D, 2021MNRAS.502L..50S}, supernova and kilonova events in the AGN disc \citep{2021MNRAS.507..156G, 2021ApJ...906L..11Z}, and tidal disruption events \citep{1988Natur.333..523R, 2019ApJ...881..113C}.

We are then interested in studying the conditions where the flares of the emerging cocoon exceed the AGN typical variability within the shortest possible time lags, as these solutions could be better identified as GW counterparts. In the following subsections, we explore the parameter space of the present model to study the behaviour of the amplitude, time lag, and duration of the predicted flares. We define as distinguishable those flare solutions representing more than 50\% of the hosting AGN, or equivalently, magnitude differences of $\lvert \Delta m \rvert \gtrsim 0.5$ mag in optical bands\footnote{AGN intrinsic variability is typically observed with $\lvert \Delta m \rvert$ of a few tenths of magnitudes in optical bands \citep{2012ApJ...753..106M, 2017MNRAS.470.4112G, 2018MNRAS.476.2501A}.}.

\subsection{Fractional excess and time delay}
\label{ssec:detectability}

   \begin{figure*}
   \centering
   \includegraphics[width=\hsize]{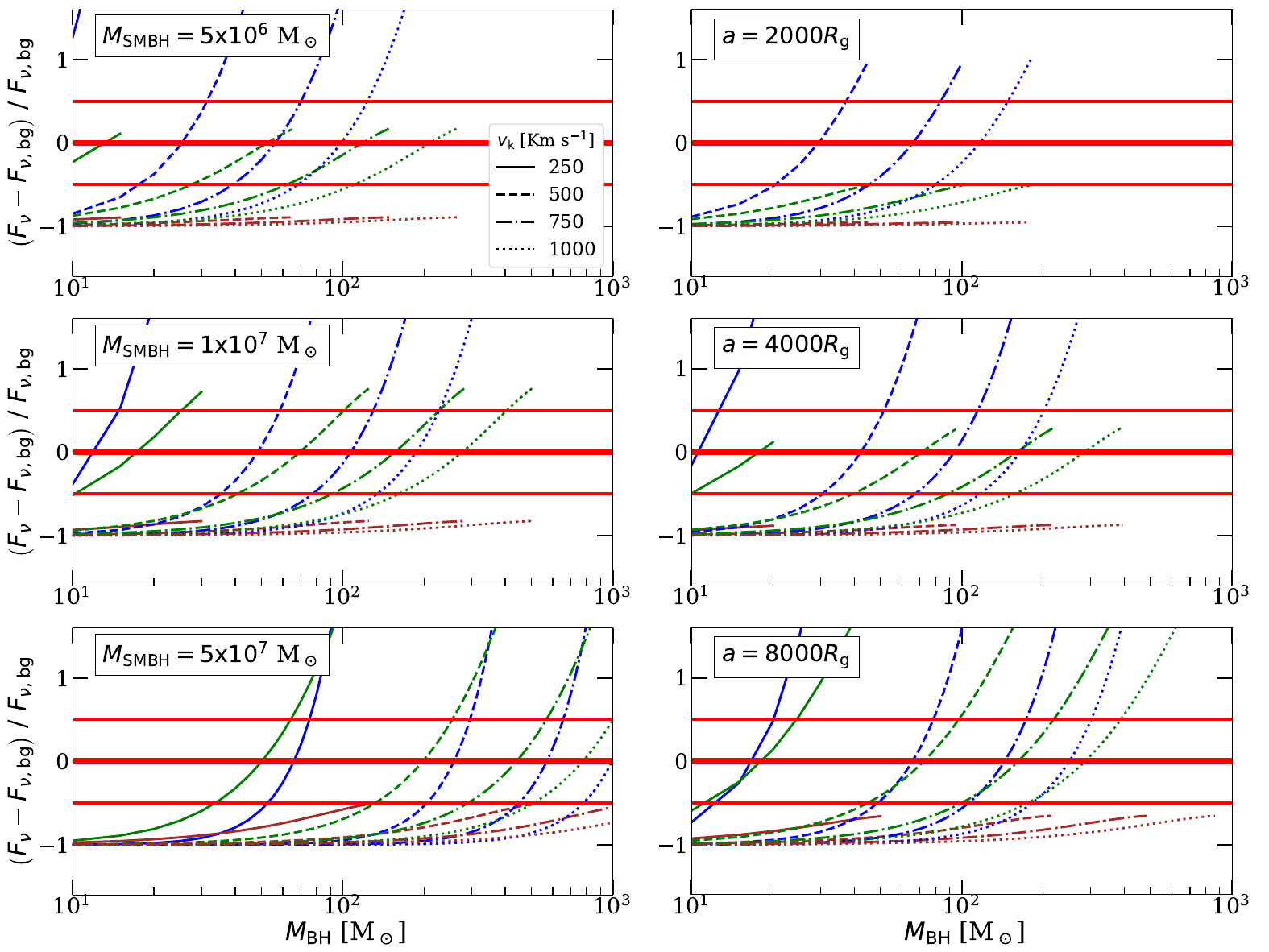}
      \caption{
Fractional excess of the emerging cocoon emission relative to the AGN background (see the text) as a function of the mass of the BH remnant.
As indicated, the curves with different line styles correspond to
different kick velocities.
The curves in brown, green and blue correspond to 
the fluxes at NIR, optical, and extreme ultraviolet frequencies, respectively, considered in  
Figure~\ref{fig:NIR-EUV_LCsvks}. 
The curves in the left panels were calculated with different SMBH 
masses, as indicated 
using the fixed values of $\dot{m}=0.05$ and $a = 5000 R_\mathrm{g}$.
The curves in the right panels were calculated with
different distances $a$ from the SMBH as indicated with the fixed values of $\dot{m}=0.05$ and $M_\mathrm{SMBH} = 10^{7}$ M$_\odot$.}
         \label{fig:rnu}
   \end{figure*}

   \begin{figure*}
   \centering
   \includegraphics[width=\hsize]{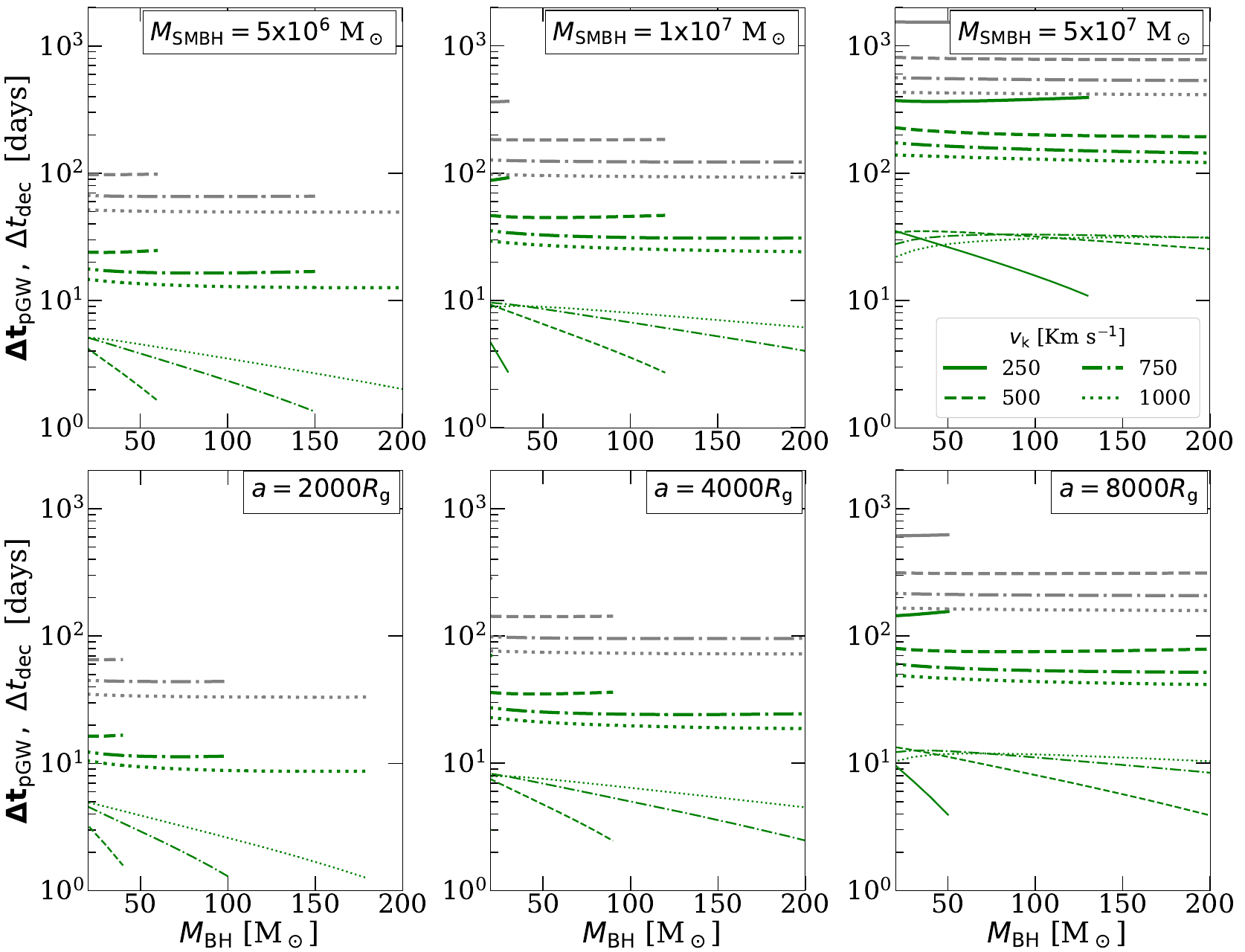}
      \caption{
Time delay for the appearance of the flare 
after the GW event (thick curves) and its duration
(thin curves) as functions of the mass of the remnant.
The curves correspond to different BH kick
velocities, as labelled.
The curves in the upper panels were obtained 
using $\dot{m}=0.05$ and $a = 5000 R_\mathrm{g}$ and different masses for the SMBH, as indicated. 
The curves in the lower panels were calculated considering 
BBH coalescences at different distances from the SMBH, as indicated, 
and using the fixed values of $\dot{m}=0.05$ and 
$M_\mathrm{SMBH} = 10^{7}$ M$_\odot$.
The source is assumed at redshift $z=0.438$.
}
         \label{fig:Atduration}
   \end{figure*}

The observed frequencies where the flare could be better
distinguished over the AGN background can vary
depending on the parameter configuration of 
the system (i.e., AGN plus the BH remnant).
Here we study the visibility of the 
emerging cocoon,
considering the fractional excess 
$r_{\nu} = (F_\nu - F_{\nu,\mathrm{bg}}) / F_{\nu,\mathrm{bg}}$
which measures the  intensity of the 
flare flux $F_\nu$ relative to 
the AGN background emission $F_{\nu,\mathrm{bg}}$.
In Figure~\ref{fig:rnu},  we display $r_{\nu}$
as a function of the mass of the remnant 
for different values of $M_\mathrm{s}$ (SMBH mass), 
$v_\mathrm{k}$ (kick velocity), and
$a$ (distance from the SMBH).
We calculate $r_\nu$ with 
$F_\nu$ at $t_\mathrm{max}$ (see 
subsection \ref{subsec:outflow-exp}),
which corresponds to the maximum bolometric luminosity 
of the flare.
We consider BH kick velocities
within the range of [250 - 1000] Km s$^{-1}$, and the three wavelengths
chosen for the light curves of Figure~\ref{fig:NIR-EUV_LCsvks}, representing the NIR, optical, and EUV domains. 
The end of the curves in this figure indicates 
that the condition (\ref{hdcond})  
for the production of 
an emerging cocoon
is no longer satisfied, according to the analysis of Section~\ref{sec:model}.
We indicate with the red horizontal lines the values of 
$r_\nu =$ -0.5, 0, and 0.5, which correspond to 
flares emitting 
50\%, 100\% and 150\% of the background emission.

The panels on the left in Figure~\ref{fig:rnu} 
explore the effect of different SMBH masses,
considering the fixed location of
$a=5000 R_\mathrm{g}$.
We note that in AGNs with SMBHs of $5\times10^{6}$ M$_\odot$, 
BH remnants should have masses lower than
$\sim 100$ M$_\odot$ to produce detectable flares, 
and such flares can be produced by 
remnants with masses as low as $\sim 30$ M$_\odot$.
As the mass of the SMBH increases, 
the mass of the BH remnant should be larger
to produce detectable flares. 
In an AGN with a SMBH of $5\times10^{7}$
M$_\odot$, for instance 
(left lower panel of Figure~\ref{fig:rnu}), 
remnants should be heavier than $\sim 50$M$_\odot$
to produce flares exceeding the background emission.
The right panels of Figure~\ref{fig:rnu} are obtained
with the fixed SMBH mass of $10^{7}$ M$_\odot$ and explore
the effect of different locations $a$ where the 
remnant interacts with the disc.
We note a trend where for larger values of $a$,
the fractional excess $r_\nu$ is enhanced.
We also note that fluxes calculated at 
the EUV wavelength
(blue curves) are, in general, more visible 
than their optical and NIR counterparts 
(blue and brown curves, respectively).
This can be seen by comparing the 
fractional excess $r_\nu$
among the curves corresponding to the same kick velocity (i.e., with the same line style).
We note that the EM counterpart 
can be seen at NIR provided that the BBH merger
occurs at distances of $a\gtrsim 8000 R_\mathrm{g}$,
in AGNs with SMBHs $\gtrsim 5\times10^{7}$ M$_\odot$.
Nevertheless, at those radii the Shakura \& Sunyaev disc model
employed here could be no longer valid (see Section~\ref{sec:profiles}).

In Figure~\ref{fig:Atduration}, we show the predicted time delay
$\Delta t_\mathrm{pGW}$ in the observer frame after the GW event
(thick curves)
for the appearance of 
the flare  and its duration $\Delta t_\mathrm{dec}$ (thin curves) as a function of the
mass of the remnant using the same parameter configurations
considered in Figure~\ref{fig:rnu}
(for $M_\mathrm{s}$,
$\dot{M}_\mathrm{s}$, 
$v_\mathrm{k}$, and
$a$). 
The time delay for the appearance of the flare is calculated with equation
(\ref{Atpgw}).
We take the flare duration as the elapsed time among the 
flare appearance and when the initial flux diminishes a factor of $e=2.718$.
The curves in the upper panels of Figure~\ref{fig:Atduration}
explore different SMBH masses using 
the fixed value of $a=5000 R_\mathrm{g}$, whereas the lower
panels explore different values of $a$ with the fixed  
SMBH mass of $10^{7}$ M$_\odot$.
We note that flares are faster as 
$M_\mathrm{s}$ and $a$ are shorter.
For instance, detectable flares
lasting as short as $\sim 2-5$ days can be produced at radii of $\sim 2000 R_\mathrm{g}$,
(see left-lower panel in Figure~\ref{fig:Atduration}).
On the other hand, relatively prominent flares like the
ones on the left-lower panel of Figure~\ref{fig:rnu},
are relatively slower. 
This strong flares are
produced by remnants 
interacting at $a \gtrsim 8000 R_\mathrm{g}$ and 
take $\sim$[100-300] days ([500-1000] days) to appear
for $z_0=5 h_\mathrm{d}$  ($z_0=20 h_\mathrm{d}$),
having durations of $\sim$ [10-30] days.
Moreover,  we note that $\Delta t_\mathrm{pGW}$
and $\Delta t_\mathrm{dec}$ increase for larger
$M_\mathrm{s}$
and $a$. 
Finally, we note that $\Delta t_\mathrm{pGW}$
do not vary significantly with the mass of the remnant,
and $\Delta t_\mathrm{dec}$
decreases as the mass of the remnant is larger.


\subsection{Optical light-curve and detectability}

\begin{figure*}
   \centering
   \includegraphics[width=\hsize]{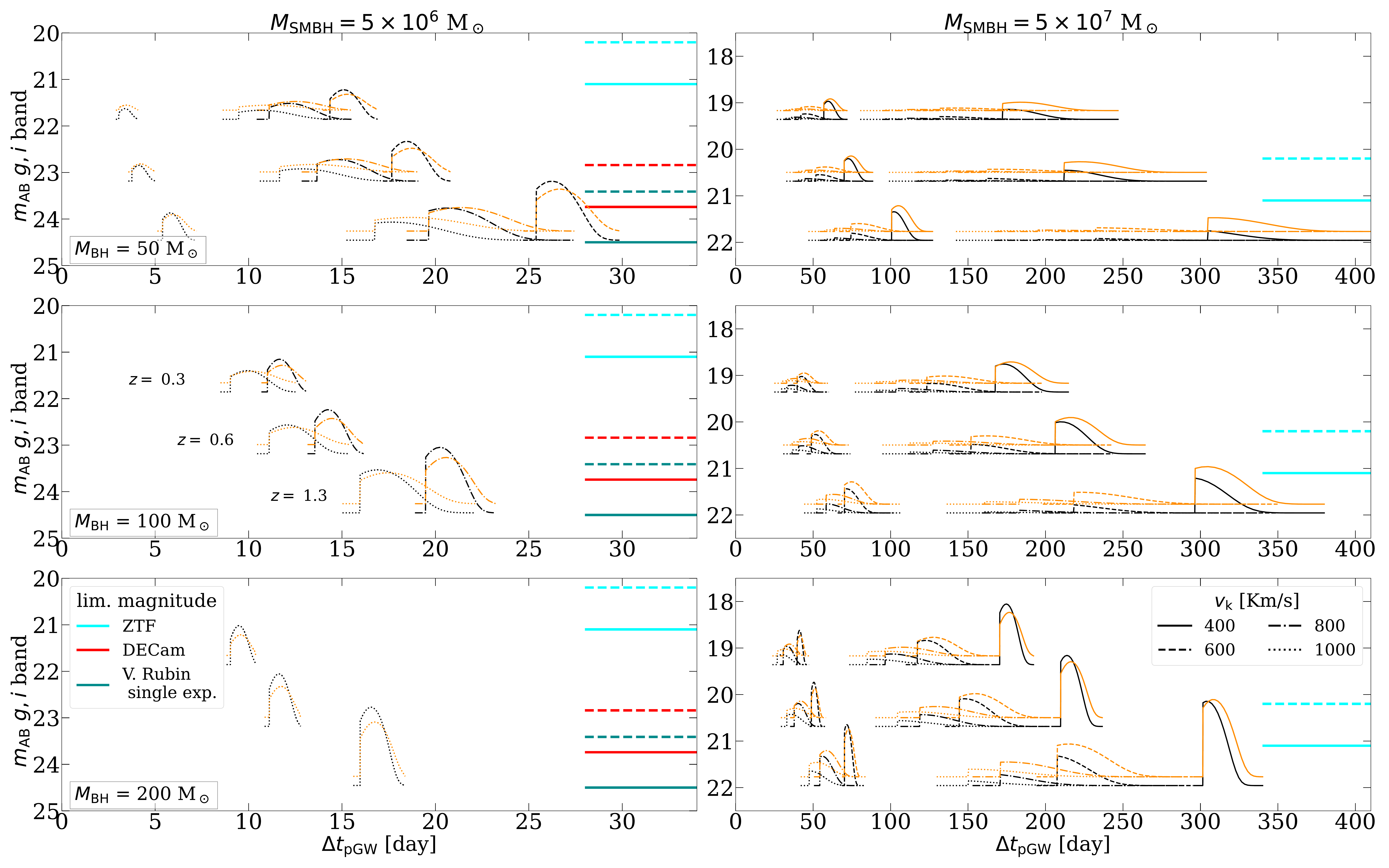}
\caption{
LC profiles in the $g$ and $i$ optical bands (black and orange curves, respectively) of the EM counterpart discussed here at different redshifts of the source.
We overplot the estimated limiting magnitudes of the ZTF \citep{2019PASP..131a8002B}, DECam (\citealt{bom2023kn}), and Vera C. Rubin (\citealt{2019ApJ...873..111I}) instruments for the $g$ and $i$ bands (solid and dashed horizontal lines, respectively).
The curves in the left and right panels correspond to AGNs with SMBHs of 
 $5\times 10^{6}$ and $5\times 10^{7}$ M$_\odot$, respectively, whereas the curves in the upper, middle, and lower panels consider remnants of $M_\mathrm{BH}=50$, $100$, and 200 M$_\odot$, respectively. In each panel, the upper, middle, and lower curves correspond to sources at redshifts of $z=0.3$, $0.6$, and $1.3$, respectively.
We explore LC solutions of remnant-disc interactions occurring at $a=1500$ and $4000$ $R_\mathrm{g}$. These solutions correspond to the left and right curve families, at a given redshift.
In the panels with one curve family at a given redshift, the LCs correspond to $a=4000 R_\mathrm{g}$, since the flare cannot be produced at $a=1500 R_\mathrm{g}$ within the present analysis (see the condition~\ref{hdcond} in Section~\ref{sec:model}).
}
         \label{fig:zsLCs}
   \end{figure*}

The flares produced by the emerging cocoon can be comparable to or exceed the emission of the hosting AGN at optical bands and last for a few days to a few weeks within the parameter space of interest (see Figures~\ref{fig:rnu}-\ref{fig:Atduration}). Thus, the flares predicted in this work can be detected by optical time-domain surveys capable of capturing transients of a few days or longer, such as the ongoing Zwicky Transient Facility (ZTF; \citealt{2019PASP..131a8002B}), the forthcoming Vera C. Rubin Observatory (LSST; \citealt{2019ApJ...873..111I}), and possible systematic searches using DECam \citep{bom2023kn, morgan2020constraints}.
The detectability of the flares also depends on the distance of the source as well as the sensitivity of the observing instrument. In Figure~\ref{fig:zsLCs}, we present examples
of the flare optical LCs 
exploring different redshifts to the source and compare them with the limiting magnitudes in the $g$ and $i$ bands of ZTF, DECam , and LSST. 
For DECam search we assumed a 60s exposure $5$-$\sigma$ detection on a dark night \citep{bom2023kn}, while for LSST we assumed the default 30s ($2\times15\mathrm{s}$) exposures \footnote{The Rubin default expossure and limiting magnitudes can be consulted in \url{https://www.lsst.org/scientists/keynumbers}}\citep{2019ApJ...873..111I}.

The LCs illustrated in the plot array of Figure~\ref{fig:zsLCs} are obtained using
the fiducial AGN accretion rate of $\dot{M}/\dot{M}_\mathrm{Edd}=0.05$,
and consider SMBHs of $5\times10^{6}$ and $5\times10^{7}$ M$_\odot$
(left and right columns, respectively).
All curves are calculated assuming mergers
occurring at $z_0 = 5h_\mathrm{d}$ form the disc mid-plane
and we explore different redshifts for the source, radial location 
of the merger $a$, and masses for the merger remnant, as indicated.
For a given redshift, we set the luminosity distance of the source $D_\mathrm{L}$ with a $\Lambda$CDM cosmology constrained by data from Planck 2018 \citep{2020A&A...641A...6P}.
To convert the flux given by the present model to bandpass AB magnitudes, we employ the \texttt{Python} library \texttt{Speclite}\footnote{
\href{https://speclite.readthedocs.io/en/latest/index.html}{https://speclite.readthedocs.io/en/latest/index.html}}
choosing the $g$ and $i$ filters (black and orange curves in Figure~\ref{fig:zsLCs}, respectively)
of the Sloan Digital Sky Survey (SDSS, \citealt{sdss_main}).

For the parameter set of Figure~\ref{fig:zsLCs},
the flares generally produce the largest magnitude variations $\lvert \Delta m \rvert$
in the $g$-band.
In this band, flares from AGNs with SMBHs of 
$\sim5\times10^{6}$ M$_\odot$ can be detected up to $z\sim 0.25$ by ZTF,
up to $z\sim 0.9$ by DECam, and
up to $z\sim 1.3$ by LSST.
These flares appear$\sim$[10-25] days after the GW event, last about 5 days, and require remnants with kicks exceeding 600 km s$^{-1}$.
Alternatively, in AGNs with SMBHs of $\sim5\times10^{7}$ M$_\odot$, the flares can be detected by ZTF up to $z\sim 0.8$, up to $z\sim 2$ by DECam,
and up to $z\sim 3$ by LSST.
These flares have onsets ranging within $\sim$[50-300] days after the GW event, last about $\sim$[20-40] days,
and their amplitude increases as the mass of the remnant is larger and the kick velocity smaller.

The emission of the emerging cocoon is calculated using 
a photon diffusion model of a plasma in homologous expansion
(see Section~\ref{subsec:outflow-exp}), a model typically employed to interpret LCs of supernovae.
Therefore, this EM counterpart evolves qualitatively 
similar to the emission of supernovae during their most intense phase.
However, it can be distinguished from supernovae since the LC of the latter typically lasts
from $\sim 40$ days to a few months \citep{1990RPPh...53.1467W, 2009ApJ...703.2205K},
whereas the emission of the
flare discussed here lasts from 
a few days up to $\sim 40$ days, accordingly.

The lag (relative to the GW event) of the flare ranges from weeks to years
(see Figures~\ref{fig:Atduration} and \ref{fig:zsLCs}). 
The flares with the shortest time lags and
at the same time magnitude variations $\lvert \Delta m \rvert \gtrsim 0.5$, which we consider the ones that could be better associated to an observed GW event,
are produced in AGNs with SMBH  
$\lesssim 5 \times 10^{6}$ M$_\odot$, by remnants with masses $\lesssim 100$ M$_\odot$, from mergers occurring at distances $a\gtrsim 4000 R_\mathrm{g}$ from the central SMBH.

\section{Summary and discussion}
\label{sec:summary}

Active galactic nuclei (AGNs) have been proposed as plausible sites hosting a sizable fraction of the
BBH mergers producing the GW events detected by the LIGO-Virgo-Kagra (LVK) experiment \citep{2022MNRAS.517.5827F}. 
Previous analyses suggest that such GW events could be accompanied
by EM counterparts due to the interaction of the merger remnant
with the AGN disc  gas \citep{2017MNRAS.464..946S,
2019ApJ...884L..50M}.
In this paper, we work out a new astrophysical scenario leading to thermal
flares that result from the interaction of a BBH merger remnant with
an AGN thin disc.
The proposed scenario is based on the following considerations.
\begin{itemize}
\item
A recoiling and highly spinning BH is formed
due to a second or higher generation of BBH merger taking place 
at a
few thousand Schwartzchild radii from the SMBH and outside the disc.
\item 
 The kicked BH enters the dense region of the disc,
 accretes material, and launches relativistic jets
 that propagate quasi-parallel to the disc plane.
\item
One of the jet cocoons created within the disc, emerges and expands outside
the disc on the observer's side. 
We then calculate the emission of photons that diffuse and emanate from the
surface of the
emerging cocoon.
\end{itemize}

Given the conditions of the cocoon when breaking out the disc, we estimate its emission using a photon diffusion model typically employed to describe LCs of supernovae
\citep{1996snih.book.....A,2012ApJ...746..121C}.
The emerging cocoon mainly emits at  
optical and EUV wavelengths
and its LC profiles exhibit in general the following features 
(see Figures~\ref{fig:rnu},
\ref{fig:Atduration}, and \ref{fig:zsLCs}):
\begin{enumerate}
\item
The larger the mass and accretion rate of the SMBH, the
larger the time delay (after the GW) and duration of the flare.
\item
The smaller the remnant's mass, the larger the duration of the flare.
\item
The larger the kick velocity, the larger the flare duration.
\end{enumerate}

In AGNs with SMBHs of $\sim 10^{6-7}$M$_\odot$,
remnants with kick velocities in the range of 
$\sim$[300 - 1000] Km s$^{-1}$ can drive flares 
comparable to or exceeding the emission of the hosting AGN.
Such flares exhibit durations
and periods of appearance 
in different time scales, depending
on the parameter configuration of the system (see Figures~\ref{fig:rnu} and \ref{fig:Atduration}).
For instance, in AGNs with SMBHs of 
$\sim 5\times10^{6}$ M$_\odot$, remnants
with masses $\lesssim$ 100 M$_\odot$
interacting with the disc at $\sim$5000 $R_\mathrm{g}$ from the SMBH can
produce flares representing more than 100\% of the AGN 
background emission.  
These flares are relatively fast, appearing
within $\sim$[10-100] days after the GW and lasting for $\sim$[1-5] days.
Alternatively, remnants with masses $\gtrsim$ 100 M$_\odot$ can
produce flares exceeding  200\% of the background emission  
when interacting with the disc at distances $\gtrsim 5000 R_\mathrm{g}$ from SMBHs of $\sim 5\times 10^{7}$ M$_\odot$.
These bright flares can appear within [100-1000] days after 
the GW and last about [20-40] days
(see Figures \ref{fig:rnu} and \ref{fig:Atduration}).

AGNs typically exhibit stochastic variability in optical bands
with $|\Delta m|$ of a few tenths of magnitudes in timescales from days to years
\citep{2012ApJ...753..106M,2017MNRAS.470.4112G,2018MNRAS.476.2501A}.
Thus, among the possivel flare profiles derived here, we suggest that those 
with the shortest time lags and magnitude variations $|\Delta m| \gtrsim 0.5$
are the ones that could be better associated with a GW event.
Such flares are produced in AGNs with SMBHs of masses $\lesssim 5 \times 10^{6}$ M$_\odot$, by remnants with masses $\lesssim 100$ M$_\odot$, from mergers occurring at distances $\gtrsim 4000 R_\mathrm{g}$ from the central SMBH (see Figures~\ref{fig:rnu}, \ref{fig:Atduration}, and \ref{fig:zsLCs}).
These flares appear within $\sim$[10-100] days after the GW event, lasting about 5 days, and require remnants with kicks exceeding 600 km s$^{-1}$.
Flares in AGNs with SMBHs of such size can be detected 
up to $z\sim 0.25$ by ZTF,
up to $z\sim 0.9$ by DECam, and
up to $z\sim 1.3$ by LSST (see Figure~\ref{fig:zsLCs}).

The present analysis is appropriate for
BBH merger remnants with masses $\gtrsim 50$
M$_\odot$, consistent with the situation 
of a second or higher-generation merger in a hierarchical sequence.
Motivated by the fact that recoils from previous coalescence can perturb the alignment of the  BHs from the disc plane,
we assume the BH remnant of the present analysis was born outside the disc.
Nevertheless, the  present model could
also be applied to coalescences
occurring within the disc,
provided that the pre-merger cavity 
\citep{2021ApJ...916..111K} is much smaller than the disc thickness.

There have been reported so far over seven 
optical flares candidates associated with BBH GW events in AGNs 
\citep{2020PhRvL.124y1102G,2023ApJ...942...99G}.
Although none of them are yet confirmed,
their theoretical interpretation is timely
to favour or disfavour such associations.
In a forthcoming work, we will investigate whether
the current EM counterparts candidates can be
explained in terms of the emission scenario discussed here.

The present emission scenario can also be employed to interpret
optical-EUV flares from AGNs with no associated GW event.
 Such emissions would correspond to flares produced by BHs orbiting around the AGN central engine
that eventually threads the disc and emerges on the observer's side.
In this case, there would be no observational constraint for the flare starting time, contrary to the case of EM flares with an associated GW signal.

The thermal emission model discussed in this paper relies on an
jet cocoon that emerges as a 
non-relativistic and quasi-spherical outflow of matter outside the AGN disc.
Here we propose that such emerging cocoons are more likely driven by jets propagating within and quasi-parallel to the disc rather than jets propagating in the perpendicular direction.
The latter case would lead to jetted and faster emerging cocoons for which the spherical expansion approximation assumed here would not apply. 
It remains as subject of further investigation 
to assess how frequently BBH remnants produce relativistic jets
propagating quasi-parallel to the disc.

Clear limitations of the present analytic model are the assumed 
morphology for 
the jet cocoon and the disc structure. 
Given the ejection of relativistic jets within the disc, 
the jet cocoons can be bent by ram pressure due to the motion of the BH remnant.
In addition, the disc vertical density stratification (not
considered in the present analysis) is expected to influence the jet  morphology and dynamics 
and hence the properties of the emerging cocoon.
Such effects could be accounted with
3D magneto-hydrodynamical simulations of the problem discussed here.
Thus, if BBH mergers indeed produce measurable EM 
counterparts, the predictions of the present analytic model can be
employed to benchmark limit cases of more realistic calculations
seeking to reconstruct the astrophysical processes 
leading to multimessenger emission from BBH mergers.

\section*{Acknowledgements}
We are thankful to the anonymous reviewer for providing valuable
critics that improved the quality of this work. 
JCRR acknowledges support from Rio de Janeiro State Funding Agency FAPERJ, grant E-26/205.635/2022.
RN acknowledges support from the Fundação de Amparo à pesquisa do Estado de São Paulo (FAPESP) for supporting this research under grant 2022/10460-8. R.N. acknowledges a Bolsa de Produtividade from Conselho Nacional de Desenvolvimento Cient\'ifico e Tecnol\'ogico.
Clecio Bom acknowledges the financial support from CNPq (316072/2021-4) and from FAPERJ (grants 201.456/2022 and 210.330/2022) and the FINEP contract 01.22.0505.00 (ref. 1891/22). The authors made use of Sci-Mind servers machines developed by the CBPF AI LAB team and would like to thank P. Russano and M. Portes de Albuquerque for all the support in infrastructure matters.
CRB and JCRR would like to thank James Annis for useful discussions at the first steps of this work.
JCRR thanks Gabriel Teixeira for useful suggestions in employing the
Python package Speclite. 

\section*{Data Availability}
The current results are given by analytic expressions and therefore, reproducible. The data from the plots will be shared upon reasonable request to the corresponding author.
 


\bibliographystyle{mnras}
\bibliography{refs} 




\appendix

\section{Displacement of a kicked BH from the disc mid-plane}
\label{app:kzmax}

Here we compute the maximum distance to the plane of an AGN disc attained by
a kicked BH remnant originated in a BBH coalescence within such disc.
In the following analysis, we neglect effects of dynamical friction
on the motion of the BBH as well as on the merger remnant.

We consider a BBH system orbiting at a distance $R_0\gg 100 R_\mathrm{g}=GM_\mathrm{s}/c^2$
from an SMBH of mass $M_\mathrm{s}$. 
At these location, general relativistic effects on the orbit of the binary system due to the gravitational field of the SMBH are negligible.
We consider for simplicity, a BBH system in  circular orbit around the SMBH.
The magnitude of the velocity of the BBH system
in the frame of the SMBH is then 
$v_\mathrm{BBH}=\sqrt{GM_\mathrm{s}/R_0}\ll c$.
At coalescence, a BH remnant of mass $M_\bullet$ is born 
with a kick velocity of typical magnitude 
$v_\mathrm{k}\sim$[100 - 1500] Km s$^{-1}\ll c$ 
in the binary co-moving frame, 
and we parameterise the kick direction
by the polar and azimuthal angles
$\theta_\mathrm{k}$ and $\varphi_\mathrm{k}$, respectively.
We then estimate the initial velocity of the remnant in the SMBH frame
as the vectorial summation of BBH and kick velocities, 
$\bar{v}_0 = \bar{v}_\mathrm{BBH} + \bar{v}_\mathrm{k}$,
and neglect special relativistic effects to 
describe the remnant's orbit.

Consider a frame where the SMBH is at the origin,
the mid-plane of the AGN disc coincides with the $x$-$y$ plane, and
the coalescence occurs at $\bar{x}_0 = (R_0,0,0)$.
In this SMBH frame, the Cartesian components of remnant's 
initial velocity $\bar{v}_0$ are:
\begin{align}
\label{v0x}
&v_{0,x} = v_\mathrm{k}\sin(\theta_\mathrm{k})\cos(\varphi_\mathrm{k}),\\
\label{v0y}
&v_{0,y} = v_\mathrm{k}\sin(\theta_\mathrm{k})\sin(\varphi_\mathrm{k})+
\sqrt{\frac{G M_\mathrm{S}}{R_0}},\\
\label{v0z}
&v_{0,z} = v_\mathrm{k}\cos(\theta_\mathrm{k}),
\end{align}
where
$\theta_\mathrm{k}$ and $\varphi_\mathrm{k}$ are measured
with respect to the $z$ and $x$ axes, respectively,  of the SMBH frame.

The orbit of the remnant lies in the plane
defined by the the vectors $\bar{x}_0$ and
$\bar{v}_0$. In such plane, the trajectory of the remnant is
described by
the usual orbit equation \citep{fetter2003theoretical}.
Thus, the $z$ coordinate
of the remnant in the SMBH frame
can be parameterised as:
\begin{align}
\nonumber
&z_\mathrm{ko} (\phi,v_\mathrm{k}, \theta_\mathrm{k}, \varphi_\mathrm{k}) =\\
&\frac{(v_{0,z}/v_\mathrm{0,y})}{\sqrt{1+(v_{0,z}/v_\mathrm{0,y})^2}}
\frac{\ell^2}{M_\bullet^2 G M_\mathrm{S} }
\frac{\sin(\phi)}{1+e\cos(\phi+\delta)},
\label{zko}
\end{align}
where $\phi$ is the angle that the remnant makes with
the periapsis of its orbit (in the orbit plane),
$\delta$ and $e$ are the phase and eccentricity of the
orbit given by 
\begin{align}
\delta & = \arccos
\left\{
\frac{1}{e}\left(\frac{\ell^2}{M_\bullet^2 G M_\mathrm{s} R_0} -1 \right)
\right\},\\
\label{ecc}
e & = \sqrt{1+\frac{2E\ell^2}{M_\bullet^3(GM_S)^2}},
\end{align}
respectively, and $\bar{\ell}$ is the remnant's angular momentum
with respect to the SMBH, which magnitude is 
\begin{equation}
\lvert \bar{\ell} \rvert=
M_\bullet R_0 \sqrt{v_{0,y}^2 + v_{0,z}^2},
\end{equation}
being 
$E = M_\bullet v_0^2/2  - GM_\bullet M_\mathrm{S}/R_0$ 
the total energy of the remnant
which is assumed as conserved in the 
present analysis.

The remnant can not be retained by the SMBH 
gravitational potential when its initial velocitiy
$v_0 = (v_\mathrm{x,0}^2+v_\mathrm{y,0}^2+v_\mathrm{z,0}^2)^{1/2}$
(which components are given by equations
\ref{v0x} - \ref{v0z})
is larger than the escape velocity

\begin{equation}
v_\mathrm{esc} = \sqrt{\frac{2GM_\mathrm{s}}{R_0}}.
\label{vesc}
\end{equation} 
Combining equations (\ref{v0x})-(\ref{v0z}), and (\ref{vesc}), the condition for the remnant's escape 
can be written as
\begin{equation}
\frac{v_0}{v_\mathrm{esc}}=
\sqrt{\frac{1}{2} + 
\sqrt{2}\sin(\theta_\mathrm{k})\sin(\varphi_\mathrm{k})
\left(
\frac{v_\mathrm{k}}{v_\mathrm{esc}}+
\right)
+
\left(
\frac{v_\mathrm{k}}{v_\mathrm{esc}}
\right)^2
}
\,\, \geqslant
1,
\end{equation}
which occurs when the remnant's kick
velocity is
\begin{equation}
v_\mathrm{k}
\geqslant
\omega_\mathrm{k} v_\mathrm{esc},
\end{equation}
where
\begin{equation}
\label{wk}
\omega_\mathrm{k} (\theta_\mathrm{k}, \varphi_\mathrm{k}) =
\frac{\sqrt{2}}{2}
\left(
\sqrt{
1+
\left[\sin(\theta_\mathrm{k})
\sin(\varphi_\mathrm{k})
\right]^2
} 
-
\sin(\theta_\mathrm{k})
\sin(\varphi_\mathrm{k})
\right).
\end{equation} 
When the product 
$\sin(\theta_\mathrm{k})
\sin(\varphi_\mathrm{k})
$
takes the lower and upper values of [-1,1], then 
$\omega_\mathrm{k} = 
[1+\sqrt{2}/2, 1-\sqrt{2}/2] \approx [1.707, 0.292]$, respectively.
Given the magnitude and direction of 
the kick velocity, we note that the remnant
escapes the AGN potential when the merger occurs 
at radii
\begin{equation}
\label{Resc}
\frac{R_0}{R_\mathrm{g}}
\geqslant 
2
\left(
\frac{c \omega_\mathrm{k}}{v_\mathrm{k}}
\right)^2
= 7200
\left(
\frac{\omega_\mathrm{k}}{0.3}
\right)^2
\left(
\frac{v_\mathrm{k}}{1500\,\mathrm{Km}\,\mathrm{s}^{-1}}
\right)^{-2}
.
\end{equation}

Alternatively, 
if $v_0<v_\mathrm{esc}$
then $e<1$ 
(see equation \ref{ecc}),
and thus the remanant follows an elliptical orbit.
In this case, the
vertical coordinate of the orbit
given by equation (\ref{zko}) 
has two extreme values, namely the
maximum displacements above and bellow the disc plane.
These extreme points occurs at 
\begin{equation}
\phi^{\pm} = 
\pm \arccos\left(1 - \frac{\ell^2}{M_\bullet^2 G M_\mathrm{s} R_0} \right).
\end{equation}

To investigate how much the remnant departs
from the AGN plane we consider the average of the two maximum displacements
\begin{equation}
\label{barze}
\bar{z}_\mathrm{e} = \frac{1}{2}\left[z_\mathrm{ko}(\phi^{+})+z_\mathrm{ko}(\phi^{-})\right],
\end{equation} 
and compare it with the disc semi-height $h_\mathrm{d}$, the latter calculated
through the SS disc model.
We also compare the time that the remnant spend within the disc
\begin{equation}
t_\mathrm{d} =
\frac{4h_\mathrm{d}}{v_\mathrm{k}\cos(\theta_\mathrm{k})},
\end{equation}
with its orbital period
\begin{equation}
T_\mathrm{orb} = 2\pi\sqrt{\frac{a_\mathrm{ax}^3}{G M_\mathrm{s}}},
\end{equation}
where
\begin{equation}
a_\mathrm{ax} = \frac{\ell^2}{M_\bullet^2 GM_\mathrm{s}}
\left[
\frac{1}{1-e^2},
\right]
\end{equation}
is the semi-major axis of the remnant's orbit.

   \begin{figure}
   \centering
   \includegraphics[width=\hsize]{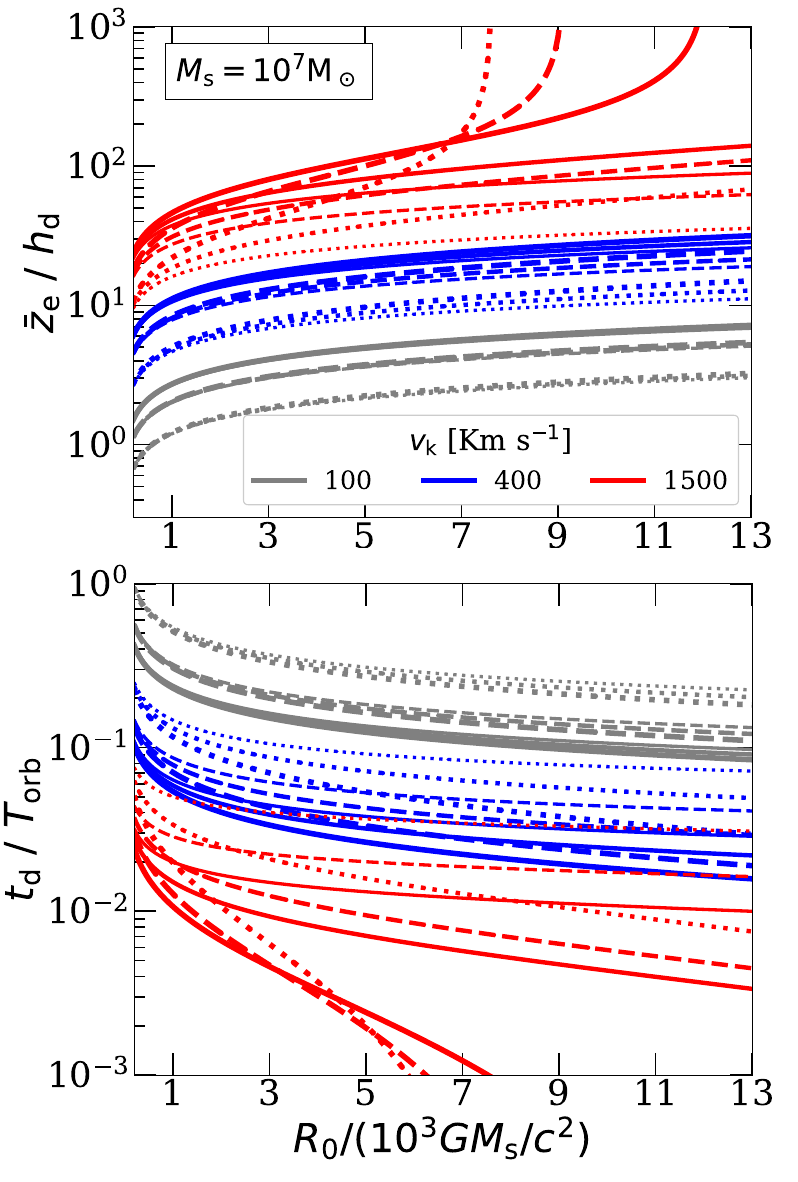}
      \caption{
Upper: Maximum vertical displacement $\bar{z}_\mathrm{e}$ attained by a remnant of a BBH merger 
occurring at a distance $R_0$ from an SMBH of $10^{7}$ M$_\odot$. 
Curves of different colours correspond to different kick velocities, as indicated.
Curves with different line styles correspond to different polar angles $\theta_\mathrm{k}$ 
of the kick direction, whereas curves with different thickness are obtained varying the azimuthal angle $\varphi_\mathrm{k}$ (see the text).
Lower: Time that the kicked remnant spends within the disc, relative to its orbital period (see the text). The different curves correspond to the parameter set of the upper panel.
}
         \label{fig:zdisp}
   \end{figure}

In Figure~\ref{fig:zdisp}, we display $\bar{z}_\mathrm{e}/h_\mathrm{d}$ 
and $t_\mathrm{d}/T_\mathrm{orb}$ as a function of the
normalised radius $R_0/R_g$
($R_\mathrm{g} = GM_\mathrm{S}/c^2$) 
for the case of an SMBH of 
$M_\mathrm{S}=10^7$ M$_\odot$.
We note that such curves are not significantly sensitive to the SMBH mass, 
and we obtain very similar curves when considering 
$M_\mathrm{S}=10^6$ and $10^{8}$  M$_\odot$.
Curves in grey, blue, and red colours correspond to remnants with 
kick velocities of 100, 400, and 1500 Km s$^{-1}$, respectively.
Curves with solid, dashed, and dotted line-styles,
are obtained using 
$\theta_\mathrm{k} = [40^\circ , 55^\circ, 70^\circ]$, respectively,
whereas thin, middle, and thick curves, correspond to
$\varphi_\mathrm{k} = [-80^\circ, 0^\circ, 80^\circ]$, respectively\footnote{
Adopting the complementary angles 
$\theta_\mathrm{k} = [110^\circ, 125^\circ, 140^\circ]$, and
$\phi_\mathrm{k} = [100^\circ, 180^\circ, 260^\circ]$
one obtains identical results due the symmetry
of the problem.}. 

For mergers occurring at $R_0\lesssim 10^4 R_\mathrm{g}$, 
we note the following features.
Remnants depart at most $\lesssim 25
h_\mathrm{d}$ from the disc plane, if they receive kicks of 
$v_\mathrm{k}\lesssim 400$ Km s$^{-1}$.
If on the other hand, remnants are born with kicks of $\gtrsim 1500$ Km s$^{-1}$,
they depart $\gtrsim 100 h_\mathrm{d}$ from the disc plane, having 
the chance to escape the AGN potential depending on the kick direction.
For instance, the diverging curves in the upper panel of
Figure~\ref{fig:zdisp} indicate the radius beyond which 
the function $\bar{z}_\mathrm{e}$ is no longer defined, namely 
where the orbit followed by the remnant is unbounded.
These escaping cases correspond to remnants with 
$v_\mathrm{k}=1500$ Km s$^{-1}$,
$\varphi_\mathrm{k} = 80^\circ$, and $\theta_\mathrm{k}= [40^\circ$, $55^\circ$,  $70^\circ$].
With this set of parameters one obtains 
through equations (\ref{wk}-\ref{Resc})
$\omega_\mathrm{k} =$ 0.389, 0.338, 0.309, 
and the minimum escaping radii of
 12121, 9143, and 7641 $R_\mathrm{g}$, 
respectively.
From the lower panel of Figure~\ref{fig:zdisp},
we note that remnants with $v_\mathrm{k}\sim$[100-400] Km s$^{-1}$
spend among $\sim $50\% and 5\% of their orbit time within the disc, 
whereas remnants with $v_\mathrm{k}\gtrsim 1500$ Km s$^{-1}$ spend $\sim$ 5\% or less
of their time within the disc.

\bsp	
\label{lastpage}
\end{document}